\documentclass[11pt,a4paper]{article}
\usepackage{amssymb}

\usepackage{amsmath}

\newcounter{resultnum}[section]

\newcounter{conclusionnum}[section]

\newcounter{conditionnum}[section]

\newcounter{conjecturenum}[section]

\newcounter{examplenum}[section]

\newcounter{exercisenum}[section]

\newcounter{lemmanum}[section]

\newcounter{notationnum}[section]

\newcounter{theoremnum}[section]

\newcounter{definitionnum}[section]

\newcounter{corollarynum}[section]

\newcounter{remarknum}[section]

\newcounter{propositionnum}[section]

\newcounter{acknowledgementnum}[section]

\newcounter{algorithmnum}[section]

\newcounter{axiomnum}[section]

\newcounter{casenum}[section]

\newcounter{claimnum}[section]

\newcounter{summarynum}[section]

\newcounter{problemnum}[section]

\begin{document}

\title{Finsler and Lagrange Geometries \\
in Einstein and String Gravity }
\date{January 31, 2008}
\author{ Sergiu I. Vacaru\thanks{%
sergiu$_{-}$vacaru@yahoo.com, svacaru@fields.utoronto.ca } \\
{\quad} \\
\textsl{The Fields Institute for Research in Mathematical Science} \\
\textsl{222 College Street, 2d Floor, } \textsl{Toronto \ M5T 3J1, Canada} }
\maketitle

\begin{abstract}
We review the current status of Finsler--Lagrange geometry and
generalizations. The goal is to aid non--experts on Finsler spaces, but
physicists and geometers skilled in general relativity and particle
theories, to understand the crucial importance of such geometric methods for
applications in modern physics. We also would like to orient mathematicians
working in generalized Finsler and K\"ahler geometry and geometric mechanics
how they could perform their results in order to be accepted by the
community of ''orthodox'' physicists.

Although the bulk of former models of Finsler--Lagrange spaces where
elaborated on tangent bundles, the surprising result advocated in our works
is that such locally anisotropic structures can be modelled equivalently on
Riemann--Cartan spaces, even as exact solutions in Einstein and/or string
gravity, if nonholonomic distributions and moving frames of references are
introduced into consideration.

We also propose a canonical scheme when geometrical objects on a (pseudo)
Riemannian space are nonholonomically deformed into generalized Lagrange, or
Finsler, configurations on the same manifold. Such canonical transforms are
defined by the coefficients of a prime metric and generate target spaces as
Lagrange structures, their models of almost Hermitian/ K\"{a}hler, or
nonholonomic Riemann spaces.

Finally, we consider some classes of exact solutions in string and Einstein
gravity modelling Lagrange--Finsler structures with solitonic pp--waves and
speculate on their physical meaning.


\vskip5pt


\end{abstract}

\tableofcontents

\section{ Introduction}

The main purpose of this survey is to present an introduction to
Finsler--Lagrange geometry and the anholonomic frame method in general
relativity and gravitation. We review and discuss possible applications in
modern physics and provide alternative constructions in the language of the
geometry of nonholonomic Riemannian manifolds (enabled with nonintegrable
distributions and preferred frame structures). It will be emphasized the
approach when Finsler like structures are modelled in general relativity and
gravity theories with metric compatible connections and, in general,
nontrivial torsion.

Usually, gravity and string theory physicists may remember that Finsler
geometry is a quite ''sophisticate'' spacetime generalization when
Riemannian metrics $g_{ij}(x^{k})$ are extended to Finsler metrics $%
g_{ij}(x^{k},y^{l})$ depending both on local coordinates $x^{k}$ on a
manifold $M$ and ''velocities'' $y^{l}$ on its tangent bundle $TM.$
\footnote{%
we emphasize that Finsler geometries can be alternatively modelled if $y^{l}$
are considered as certain nonholonomic, i. e. constrained, coordinates on a
general manifold $\mathbf{V}$, not only as "velocities" or "momenta", see
further constructions in this work} Perhaps, they will say additionally that
in order to describe local anisotropies depending only on directions given
by vectors $y^{l},$ the Finsler metrics should be defined in the form $%
g_{ij}\sim \frac{\partial F^{2}}{\partial y^{i}\partial y^{j}},$ where $%
F(x^{k},\zeta y^{l})$ $=|\zeta |\ F(x^{k},y^{l}),$ for any real $\zeta \neq
0,$ is a fundamental Finsler metric function. A number of authors analyzing
possible locally anisotropic physical effects omit a rigorous study of
nonlinear connections and do not reflect on the problem of compatibility of
metric and linear connection structures. If a Riemannian geometry is
completely stated by its metric, various models of Finsler spaces and
generalizations are defined by three independent geometric objects (metric
and linear and nonlinear connections) which in certain canonical cases are
induced by a fundamental Finsler function $F(x,y).$ For models with
different metric compatibility, or non--compatibility, conditions, this is a
point of additional geometric and physical considerations, new terminology
and mathematical conventions. Finally, a lot of physicists and
mathematicians have concluded that such geometries with generic local
anisotropy are characterized by various types of connections, torsions and
curvatures which do not seem to have physical meaning in modern particle
theories but (may be?) certain Finsler like analogs of mechanical systems
and continuous media can be constructed.

There were published a few rigorous studies on perspectives of Finsler like
geometries in standard theories of gravity and particle physics (see, for
instance, Refs. \cite{1bek,1will}) but they do not analyze any physical
effects of the nonlinear connection and adapted linear connection structures
and the possibility to model Finsler like spaces as exact solutions in
Einstein and sting gravity \cite{1vsgg}). The results of such works, on
Finsler models with violations of local Lorentz symmetry and nonmetricity
fields, can be summarized in a very pessimistic form:\ both fundamental
theoretic consequences and experimental data restrict substantially the
importance for modern physics of locally anisotropic geometries elaborated
on (co) tangent bundles,\footnote{%
In result of such opinions, the Editors and referees of some top physical
journals almost stopped to accept for publication manuscripts on Finsler
gravity models. If other journals were more tolerant with such theoretical
works, they were considered to be related to certain alternative classes of
theories or to some mathematical physics problems with speculations on
geometric models and "nonstandard" physics, mechanics and some applications
to biology, sociology or seismology etc} see Introduction to monograph \cite%
{1vsgg} and article \cite{1vesnc} and reference therein for more detailed
reviews and discussions.

Why we should give a special attention to Finsler geometry and methods and
apply them in modern physics ?\ We list here a set of contr--arguments and
discus the main sources of "anti--Finsler" skepticism which (we hope) will
explain and re--move the existing unfair situation when spaces with generic
local anisotropy are not considered in standard theories of physics:

\begin{enumerate}
\item One should be emphasized that in the bulk the criticism on locally
anisotropic geometries and applications in standard physics was motivated
only for special classes of models on tangent bundles, with violation of
local Lorentz symmetry (even such works became very important in modern
physics, for instance, in relation to brane gravity \cite{1groj} and quantum
theories \cite{1kost}) and nonmetricity fields. Not all theories with
generalized Finsler metrics and connections were elaborated in this form (on
alternative approaches, see next points) and in many cases, like \cite%
{1bek,1will}, the analysis of physical consequences was performed not
following the nonlinear connection geometric formalism and a tensor calculus
adapted to nonholonomic structures which is crucial in Finsler geometry and
generalizations.

\item More recently, a group of mathematicians \cite{1bcs,1shen} developed
intensively some directions on Finsler geometry and applications following
the Chern's linear connection formalism proposed in 1948 (this connection is
with vanishing torsion but noncompatible with the metric structure). For
non--experts in geometry and physics, the works of this group, and other
authors working with generalized local Lorentz symmetries, created a false
opinion that Finsler geometry can be elaborated only on tangent bundles and
that the Chern connection is the "best" Finsler generalization of the Levi
Civita connection. A number of very important constructions with the
so--called metric compatible Cartan connection, or other canonical
connections, were moved on the second plan and forgotten. One should be
emphasized that the geometric constructions with the well known Chern or
Berwald connections can not be related to standard theories of physics
because they contain nonmetricity fields. The issue of nonmetricity was
studied in details in a number of works on metric--affine gravity, see
review \cite{1mag} and Chapter I in the collection of works \cite{1vsgg},
the last one containing a series of papers on generalized Finsler--affine
spaces. Such results are not widely accepted by physicists because of
absence of experimental evidences and theoretical complexity of geometric
constructions. Here we note that it is a quite sophisticate task to
elaborate spinor versions, supersymmetric and noncommutative generalizations
of Finsler like geometries if we work with metric noncompatible connections.

\item A non--expert in special directions of differential geometry and
geometric mechanics, may not know that beginning E. Cartan (1935) \cite%
{1cart} various models of Finsler geometry were developed alternatively by
using metric compatible connections which resulted in generalizations to the
geometry of Lagrange and Hamilton mechanics and their higher order
extensions. Such works and monographs were published by prominent schools
and authors on Finsler geometry and generalizations from Romania and Japan %
\cite%
{1ma1987,1ma,1mhl,1mhf,1mhss,1mhh,1kaw1,1kaw2,1ikeda,1takano,1watanik,1mats,1opr1,1opr2,1opp2,1oppor,1bej,1bejf}
following approaches quite different from the geometry of symplectic
mechanics and generalizations \cite{1dleon,1lib,1mard,1krup}. As a matter of
principle, all geometric constructions with the Chern and/or simplectic
connections can de redefined equivalently for metric compatible geometries,
but the philosophy, aims, mathematical formalism and physical consequences
are very different for different approaches and the particle physics
researches usually are not familiar with such results.

\item It should be noted that for a number of scientists working in Western
Countries there are less known the results on the geometry of nonholonomic
manifolds published in a series of monographs and articles by G. Vr\v
anceanu (1926), Z. Horak (1927) and others \cite{1vr1,1vr1a,1vr2,1hor}, see
historical remarks and bibliography in Refs. \cite{1bejf,1vsgg}. The
importance for modern physics of such works follows from the idea and
explicit proofs (in quite sophisticate component forms) that various types
of locally anisotropic geometries and physical interactions can be modelled
on usual Riemannian manifolds by considering nonholonomic distributions and
holonomic fibrations enabled with certain classes of special connections.

\item In our works (see, for instance, reviews and monographs \cite%
{1vfs,1vstrf,1vncsup,1vhs,1vmon1,1vstav,1vtnut,1vv,1vesnc,1vsgg}, and
references therein), we re--oriented the research on Finsler spaces and
generalizations in some directions connected to standard models of physics
and gauge, supersymmetric and noncommutative extensions of gravity. Our
basic idea was that both the Riemann--Cartan and generalized
Finsler--Lagrange geometries can be modelled in a unified manner by
corresponding geometric structures on nonholono\-mic manifolds. It was
emphasized, that prescribing a preferred nonholonomic frame structure
(equivalently, a nonintegrabie distribution with associated nonlinear
connection) on a manifold, or on a vector bundle, it is possible to work
equivalently both with the Levi Civita and the so--called canonical
distinguished connection. We provided a number of examples when Finsler like
structures and geometries can be modelled as exact solutions in Einstein and
string gravity and proved that certain geometric methods are very important,
for instance, in constructing new classes of exact solutions.
\end{enumerate}

This review work has also pedagogical scopes. We attempt to cover key
aspects and open issues of generalized Finsler--Lagrange geometry related to
a consistent incorporation of nonlinear connection formalism and moving/
deformation frame methods into the Einstein and string gravity and analogous
models of gravity, see also Refs. \cite{1vesnc,1vsgg,1ma,1bej,1rund} for
general reviews, written in the same spirit as the present one but in a more
comprehensive, or inversely, with more special purposes forms. While the
article is essentially self--contained, the emphasis is on communicating the
underlying ideas and methods and the significance of results rather than on
presenting systematic derivations and detailed proofs (these can be found in
the listed literature).

The subject of Finsler geometry and applications can be approached in
different ways. We choose one of which is deeply rooted in the well
established gravity physics and also has sufficient mathematical precision
to ensure that a physicist familiar with standard textbooks and monographs
on gravity \cite{1haw,1mtw,1wald,1sb} and string theory \cite%
{1string1,1string2,1string3} will be able without much efforts to understand
recent results and methods of the geometry of nonholonomic manifolds and
generalized Finsler--Lagrange spaces.

We shall use the terms "standard" and "nonstandard" models in geometry and
physics. In connection to Finsler geometry, we shall consider a model to be
a standard one if it contains locally anisotropic structures defined by
certain nonholonomic distributions and adapted frames of reference on a
(pseudo) Riemannian or Riemann--Cartan space (for instance, in general
relativity, Kaluza--Klein theories and low energy string gravity models).
Such constructions preserve, in general, the local Lorentz symmetry and they
are performed with metric compatible connections. The term "nonstandard"
will be used for those approaches which are related to metric
non--compatible connections and/or local Lorentz violations in Finsler
spacetimes and generalizations. Sure, any standard or nonstandard model is
rigorously formulated following certain purposes in modern geometry and
physics, geometric mechanics, biophysics, locally anisotropic thermodynamics
and stochastic and kinetic processes and classical or quantum gravity
theories. Perhaps, it will be the case to distinguish the class of "almost
standard" physical models with locally anisotropic interactions when certain
geometric objects from a (pseudo) Riemannian or Riemann--Cartan manifolds
are lifted on a (co) tangent or vector bundles and/or their supersymmetric,
non--commutative, Lie algebroid, Clifford space, quantum group ...
generalizations. There are possible various effects with "nonstandard"
corrections, for instance, violations of the local Lorentz symmetry by
quantum effects but in some classical or quantum limits such theories are
constrained to correspond to certain standard ones.

This contribution is organized as follows:

In section 2, we outline an unified approach to the geometry of nonholonomic
distributions on Riemann manifolds and Finsler--Lagrange spaces. The basic
concepts on nonholonomic manifolds and associated nonlinear connection
structures are explained and the possibility of equivalent (non) holonomic
formulations of gravity theories is analyzed.

Section 3 is devoted to nonholonomic deformations of manifolds and vector
bundles. There are reviewed the basic constructions in the geometry of
(generalized) Lagrange and Finsler spaces. A general ansatz for constructing
exact solutions, with effective Lagrange and Finsler structures, in Einstein
and string gravity, is analyzed.

In section 4, the Finsler--Lagrange geometry is formulated as a variant of
almost Hermitian and/or K\"aher geometry. We show how the Einstein gravity
can be equivalently reformulated in terms of almost Hermitian geometry with
preferred frame structure.

Section 5 is focused on explicit examples of exact solutions in Einstein and
string gravity when (generalized) Finsler--Lagrange structures are modelled
on (pseudo) Riemannian and Riemann--Cartan spaces. We analyze some classes
of Einstein metrics which can be deformed into new exact solutions
characterized additionally by Lagrange--Finsler configurations. For string
gravity, there are constructed explicit examples of locally anisotropic
configurations describing gravitational solitonic pp--waves and their
effective Lagrange spaces. We also analyze some exact solutions for
Finsler--solitonic pp--waves on Schwarzschild spaces.

Conclusions and further perspectives of Finsler geometry and new geometric
methods for modern gravity theories are considered in section 6.

Finally, we should note that our list of references is minimalist, trying to
concentrate on reviews and monographs rather than on original articles. More
complete reference lists are presented in the books \cite%
{1vsgg,1vmon1,1vstav,1ma,1mhss}. Various guides for learning, both for
experts and beginners on geometric methods and further applications in
standard and nonstandard physics, with references, are contained in \cite%
{1vsgg,1ma,1mhss,1bej,1rund}.

\section{Nonholonomic Einstein Gravity and Finsler--La\-grange Spa\-ces}

In this section we present in a unified form the Riemann--Cartan and
Finsler--Lagrange geometry. The reader is supposed to be familiar with
well--known geometrical approaches to gravity theories \cite%
{1haw,1mtw,1wald,1sb} but may not know the basic concepts on Finsler
geometry and nonholonomic manifolds. The constructions for locally
anisotropic spaces will be derived by special parametrizations of the frame,
metric and connection structures on usual manifolds, or vector bundle
spaces, as we proved in details in Refs. \cite{1vsgg,1vesnc}.

\subsection{Metric--affine, Riemann--Cartan and Einstein manifolds}

Let $V$ be a necessary smooth class manifold of dimension $\dim V=n+m,$ when
$n\geq 2$ and $m\geq 1,$ enabled with \textbf{metric}, $g=g_{\alpha \beta
}e^{\alpha }\otimes e^{\beta },$ and \textbf{linear connection}, $D=\{\Gamma
_{\ \beta \gamma }^{\alpha }\},$ structures. The coefficients of $g$ and $D$
can be computed with respect to any local \textbf{frame}, $e_{\alpha },$ and
\textbf{co--frame}, $e^{\beta },$ bases, for which $e_{\alpha }\rfloor
e^{\beta }=\delta _{\alpha }^{\beta },$ where $\rfloor $ denotes the
interior (scalar) product defined by $g$ and $\delta _{\alpha }^{\beta }$ is
the Kronecker symbol. A local system of coordinates on $V$ is denoted $%
u^{\alpha }=(x^{i},y^{a}),$ or (in brief) $u=(x,y),$ where indices run
correspondingly the values: $i,j,k...=1,2,...,n$ and $%
a,b,c,...=n+1,n+2,...n+m$ for any splitting $\alpha =(i,a),\beta =(j,b),...$
We shall also use primed, underlined, or other type indices: for instance, $%
e_{\alpha ^{\prime }}=(e_{i^{\prime }},e_{a^{\prime }})$ and $e^{\beta
^{\prime }}=(e^{j^{\prime }},e^{b^{\prime }}),$ for a different sets of
local (co) bases, or $\underline{e}_{\alpha }=e_{\underline{\alpha }%
}=\partial _{\underline{\alpha }}=\partial /\partial u^{\underline{\alpha }%
}, $ $\underline{e}_{i}=e_{\underline{i}}=\partial _{\underline{i}}=\partial
/\partial x^{\underline{i}}$ and $\underline{e}_{a}=e_{\underline{a}%
}=\partial _{\underline{a}}=\partial /\partial y^{\underline{a}}$ if we wont
to emphasize that the coefficients of geometric objects (tensors,
connections, ...) are defined with respect to a local \textbf{coordinate
basis}. For simplicity, we shall omit underlining or priming of indices and
symbols if that will not result in ambiguities. The Einstein's summation
rule on repeating ''up-low'' indices will be applied if the contrary will
not be stated.

\textbf{Frame transforms} of a local basis $e_{\alpha }$ and its dual basis $%
e^{\beta }$ are paramet\-riz\-ed in the form
\begin{equation}
e_{\alpha }=A_{\alpha }^{\ \alpha ^{\prime }}(u)e_{\alpha ^{\prime }}%
\mbox{\
and\  }e^{\beta }=A_{\ \beta ^{\prime }}^{\beta }(u)e^{\beta ^{\prime }},
\label{ft}
\end{equation}%
where the matrix $A_{\ \beta ^{\prime }}^{\beta }$ is inverse to $A_{\alpha
}^{\ \alpha ^{\prime }}.$ In general, local bases are \textbf{nonholonomic}
(equivalently, \textbf{anholonomic}, or \textbf{nonintegrable}) and satisfy
certain anholonomy conditions
\begin{equation}
e_{\alpha }e_{\beta }-e_{\beta }e_{\alpha }=W_{\alpha \beta }^{\gamma
}e_{\gamma }  \label{nhr}
\end{equation}%
with nontrivial \textbf{anholonomy coefficients} $W_{\alpha \beta }^{\gamma
}(u).$ We consider the \textbf{holonomic} frames to be defined by $W_{\alpha
\beta }^{\gamma }=0,$ which holds, for instance, if we fix a local
coordinate basis.

Let us denote the covariant derivative along a vector field $X=X^{\alpha
}e_{\alpha }$ as $D_{X}=X\rfloor D.$ One defines three fundamental geometric
objects on manifold $V:$ \textbf{nonmetricity} field,
\begin{equation}
\mathcal{Q}_{X}\doteqdot D_{X}g,  \label{nm}
\end{equation}%
\textbf{torsion},
\begin{equation}
\mathcal{T}(X,Y)\doteqdot D_{X}Y-D_{Y}X-[X,Y],  \label{ators}
\end{equation}%
and \textbf{curvature},
\begin{equation}
\mathcal{R}(X,Y)Z\doteqdot D_{X}D_{Y}Z-D_{Y}D_{X}Z-D_{[X,Y]}Z,  \label{acurv}
\end{equation}%
where the symbol ''$\doteqdot $'' states ''by definition'' and $%
[X,Y]\doteqdot XY-YX.$ With respect to fixed local bases $e_{\alpha }$ and $%
e^{\beta },$ the coefficients $\mathcal{Q}=\{Q_{\alpha \beta \gamma
}=D_{\alpha }g_{\beta \gamma }\},\mathcal{T}=\{T_{\ \beta \gamma }^{\alpha
}\}$ and $\mathcal{R}=\{R_{\ \beta \gamma \tau }^{\alpha }\}$ can be
computed by introducing $X\rightarrow e_{\alpha },Y\rightarrow e_{\beta
},Z\rightarrow e_{\gamma }$ into respective formulas (\ref{nm}), (\ref{ators}%
) and (\ref{acurv}).

In gravity theories, one uses three others important geometric objects: the
\textbf{Ricci tensor}, $Ric(D)=\{R_{\ \beta \gamma }\doteqdot R_{\ \beta
\gamma \alpha }^{\alpha }\},$ the \textbf{scalar curvature}, $R\doteqdot
g^{\alpha \beta }R_{\alpha \beta }$ ($g^{\alpha \beta }$ being the inverse
matrix to $g_{\alpha \beta }),$ and the \textbf{Einstein tensor}, $\mathcal{E%
}=\{E_{\alpha \beta }\doteqdot R_{\alpha \beta }-\frac{1}{2}g_{\alpha \beta
}R\}.$

A manifold $\ ^{ma}V$ is a \textbf{metric--affine space }if it is provided
with arbitrary two independent metric $g$ and linear connection $D$
structures and characterized by three nontrivial fundamental geometric
objects $\mathcal{Q},\mathcal{T}$ and $\mathcal{R}.$

If the metricity condition, $\mathcal{Q}=0,$ is satisfied for a given couple
$g$ and $D,$ such a manifold $\ ^{RC}V$ is called a \textbf{Riemann--Cartan
space }with nontrivial torsion $\mathcal{T}$ of $D.$

A \textbf{Riemann space} $\ ^{R}V$ is provided with a metric structure $g$
which defines a unique Levi Civita connection $\ _{\shortmid }D=\nabla ,$
which is both metric compatible, $\ _{\shortmid }\mathcal{Q}=\nabla g=0,$
and torsionless, $\ _{\shortmid }\mathcal{T}=0.$ Such a space is pseudo-
(semi-) Riemannian if locally the metric has any mixed signature $(\pm 1,
\pm 1, ..., \pm 1).$\footnote{%
mathematicians usually use the term semi--Riemannian but physicists are more
familiar with pseudo--Riemannian; we shall apply both terms on convenience}
In brief, we shall call all such spaces to be Riemannian (with necessary
signature) and denote the main geometric objects in the form $\ _{\shortmid }%
\mathcal{R}=\{\ _{\shortmid }R_{\ \beta \gamma \tau }^{\alpha }\},$ $\
_{\shortmid }Ric(\ _{\shortmid }D)=\{\ _{\shortmid }R_{\ \beta \gamma }\},\
_{\shortmid }R$ and $\ _{\shortmid }\mathcal{E}=\{\ _{\shortmid }E_{\alpha
\beta }\}.$

The \textbf{Einstein gravity theory} is constructed canonically for $\dim
^{R}V=4 $ and Minkowski signature, for instance, $(-1,+1,+1,+1).$ Various
generalizations in modern \textbf{string and/or gauge gravity} consider
Riemann, Riemann--Cartan and metric--affine spaces of higher dimensions.

The \textbf{Einstein equations }are postulated in the form
\begin{equation}
\mathcal{E}(D)\doteqdot Ric(D)-\frac{1}{2}\ g~Sc(D)=\Upsilon ,  \label{einst}
\end{equation}%
where the source $\Upsilon $ contains contributions of matter fields and
corrections from, for instance, string/brane theories of gravity. In a
physical model, the equations (\ref{einst}) have to be completed with
equations for the matter fields and torsion (for instance, in the\ \textbf{%
Einstein--Cartan theory} \cite{1mag}, one considers algebraic equations for
the torsion and its source). It should be noted here that because of
possible nonholonomic structures on a manifold $V$ (we shall call such
spaces to be locally anisotropic), see next section, the tensor $Ric(D)$ is
not symmetric and $D\left[ \mathcal{E}(D)\right] \neq 0.$ This imposes a
more sophisticate form of conservation laws on spaces with generic ''local
anisotropy'', see discussion in \cite{1vsgg} (a similar situation arises in
Lagrange mechanics \cite{1dleon,1lib,1mard,1krup,1ma} when nonholonomic
constraints modify the definition of conservation laws).

For \textbf{general relativity}, $\dim V=4$ and $D=\nabla, $ the field
equations can be written in the well--known component form%
\begin{equation}
\ _{\shortmid }E_{\alpha \beta }=\ _{\shortmid }R_{\ \beta \gamma }-\frac{1}{%
2}\ _{\shortmid }R=\Upsilon _{\alpha \beta }  \label{einstgr}
\end{equation}%
when $\nabla (\ _{\shortmid }E_{\alpha \beta })=\nabla (\Upsilon _{\alpha
\beta })=0.$ The coefficients in equations (\ref{einstgr}) are defined with
respect to arbitrary nonholomomic frame (\ref{ft}).

\subsection{Nonholonomic manifolds and adapted frame structures}

A \textbf{nonholonomic manifold} $(M,\mathcal{D})$ is a manifold $M$ of
necessary smooth class enabled with a nonholonomic distribution $\mathcal{D}%
, $ see details in Refs. \cite{1bejf,1vsgg}. Let us consider a $(n+m)$%
--dimensional manifold $\mathbf{V,}$ with $n\geq 2$ and $m\geq 1$ (for a
number of physical applications, it will be considered to model a physical
or geometric space). In a particular case, $\mathbf{V=}TM,$ with $n=m$ (i.e.
a tangent bundle), or $\mathbf{V=E}=(E,M),$ $\dim M=n,$ is a vector bundle
on $M,$ with total space $E$ (we shall use such spaces for traditional
definitions of Finsler and Lagrange spaces \cite%
{1ma1987,1ma,1mats,1bej,1rund,1bcs,1shen}). In a general case, a manifold $%
\mathbf{V}$ is provided with a local fibred structure into conventional
''horizontal'' and ''vertical'' directions defined by a nonholonomic
(nonintegrable) distribution with associated nonlinear connection
(equivalently, nonholonomic frame) structure. Such nonholonomic manifolds
will be used for modelling locally anisotropic structures in Einstein
gravity and generalizations \cite{1vhs,1vtnut,1vv,1vesnc,1vsgg}.

\subsubsection{Nonlinear connections and N--adapted frames}

We denote by $\pi ^{\top }:T\mathbf{V}\rightarrow TM$ the differential of a
map $\pi :\mathbf{V}\rightarrow M$ defined by fiber preserving morphisms of
the tangent bundles $T\mathbf{V}$ and $TM.$ The kernel of $\pi ^{\top }$ is
just the vertical subspace $v\mathbf{V}$ with a related inclusion mapping $%
i:v\mathbf{V}\rightarrow T\mathbf{V}.$ For simplicity, in this work
we restrict our considerations for a fibered
manifold  $\mathbf{V}\rightarrow M$  with constant rank $\pi.$ In
such cases, we can define connections and metrics on  $\mathbf{V}$  in usual form,
but with the aim to "mimic" Finsler and Lagrange like structures (not on usual
tangent bundles but on such nonholonomic manifolds) we shall also consider
 metrics, tensors and connections adapted to the fibered structure
 as it was elaborated in Finsler geometry (see below the concept of distinguished metric, section 2.2.2
 and distinguished connection, section 2.2.3).

A \textbf{nonlinear connection (N--connection)} $\mathbf{N}$ on a manifold $%
\mathbf{V}$ is defined by the splitting on the left of an exact sequence
\begin{equation*}
0\rightarrow v\mathbf{V}\overset{i}{\rightarrow }T\mathbf{V}\rightarrow T%
\mathbf{V}/v\mathbf{V}\rightarrow 0,
\end{equation*}%
i. e. by a morphism of submanifolds $\mathbf{N:\ \ }T\mathbf{V}\rightarrow v%
\mathbf{V}$ such that $\mathbf{N\circ i}$ is the unity in $v\mathbf{V}.$\footnote{There is a proof
 (see, for instance,  Ref. \cite{1ma}, Theorem 1.2, page 21)
  that for a vector bundle over a paracompact manifold there exist
  N--connections. In this work, we restrict our
  considerations only for fibered manifolds admitting
  N--connection and related N--adapted frame structures (see the end of this
  section).}

Locally, a N--connection is defined by its coefficients $N_{i}^{a}(u),$%
\begin{equation}
\mathbf{N}=N_{i}^{a}(u)dx^{i}\otimes \frac{\partial }{\partial y^{a}}.
\label{coeffnc}
\end{equation}%
In an equivalent form, we can say that any N--connection is defined by a
\textbf{Whitney sum} of conventional horizontal (h) space, $\left( h\mathbf{V%
}\right) ,$ and vertical (v) space, $\left( v\mathbf{V}\right) ,$
\begin{equation}
T\mathbf{V}=h\mathbf{V}\oplus v\mathbf{V}.  \label{whitney}
\end{equation}%
The sum (\ref{whitney}) states on $T\mathbf{V}$ a nonholonomic
(equivalently, anholonomic, or nonintegrable) distribution of h- and
v--space. The well known class of linear connections consists on a
particular subclass with the coefficients being linear on $y^{a},$ i.e.
\begin{equation}
N_{i}^{a}(u)=\Gamma _{bj}^{a}(x)y^{b}.  \label{lincon}
\end{equation}

The geometric objects on $\mathbf{V}$ can be defined in a form adapted to a
N--connection structure, following decompositions which are invariant under
parallel transports preserving the splitting (\ref{whitney}). In this case,
we call them to be distinguished (by the N--connection structure), i.e.
\textbf{d--objects.} For instance, a vector field $\mathbf{X}\in T\mathbf{V}$
\ is expressed
\begin{equation*}
\mathbf{X}=(hX,\ vX),\mbox{ \ or \ }\mathbf{X}=X^{\alpha }\mathbf{e}_{\alpha
}=X^{i}\mathbf{e}_{i}+X^{a}e_{a},
\end{equation*}%
where $hX=X^{i}\mathbf{e}_{i}$ and $vX=X^{a}e_{a}$ state, respectively, the
adapted to the N--connection structure horizontal (h) and vertical (v)
components of the vector. In brief, $\mathbf{X}$ is called a distinguished
vectors, \textbf{d--vector}.\footnote{%
We shall use always ''boldface'' symbols if it would be necessary to
emphasize that certain spaces and/or geometrical objects are
provided/adapted to a\ N--connection structure, or with the coefficients
computed with respect to N--adapted frames.} In a similar fashion, the
geometric objects on $\mathbf{V}$ like tensors, spinors, connections, ...
are called respectively \textbf{d--tensors, d--spinors, d--connections} if
they are adapted to the N--connection splitting (\ref{whitney}).

The \textbf{N--connection curvature} is defined as the \textbf{Neijenhuis
tensor}%
\begin{equation}
\mathbf{\Omega }(\mathbf{X,Y})\doteqdot \lbrack vX,vY]+\ v[\mathbf{X,Y}]-v[vX%
\mathbf{,Y}]-v[\mathbf{X,}vY].  \label{njht}
\end{equation}%
In local form, we have for (\ref{njht})
 $\mathbf{\Omega }=\frac{1}{2}\Omega _{ij}^{a}\ d^{i}\wedge d^{j}\otimes
\partial _{a},$ with coefficients%
\begin{equation}
\Omega _{ij}^{a}=\frac{\partial N_{i}^{a}}{\partial x^{j}}-\frac{\partial
N_{j}^{a}}{\partial x^{i}}+N_{i}^{b}\frac{\partial N_{j}^{a}}{\partial y^{b}}%
-N_{j}^{b}\frac{\partial N_{i}^{a}}{\partial y^{b}}.  \label{ncurv}
\end{equation}

Any N--connection $\mathbf{N}$ may be characterized by an associated frame
(vielbein) structure $\mathbf{e}_{\nu }=(\mathbf{e}_{i},e_{a}),$ where
\begin{equation}
\mathbf{e}_{i}=\frac{\partial }{\partial x^{i}}-N_{i}^{a}(u)\frac{\partial }{%
\partial y^{a}}\mbox{ and
}e_{a}=\frac{\partial }{\partial y^{a}},  \label{dder}
\end{equation}%
and the dual frame (coframe) structure $\mathbf{e}^{\mu }=(e^{i},\mathbf{e}%
^{a}),$ where
\begin{equation}
e^{i}=dx^{i}\mbox{ and }\mathbf{e}^{a}=dy^{a}+N_{i}^{a}(u)dx^{i},
\label{ddif}
\end{equation}%
see formulas (\ref{ft}). These vielbeins are called respectively \textbf{%
N--adapted fra\-mes and coframes.} In order to preserve a relation with the
previous denotations \cite{1vesnc,1vsgg}, we emphasize that $\mathbf{e}_{\nu
}=(\mathbf{e}_{i},e_{a})$ and $\mathbf{e}^{\mu }=(e^{i},\mathbf{e}^{a})$ are
correspondingly the former ''N--elongated'' partial derivatives $\delta
_{\nu }=\delta /\partial u^{\nu }=(\delta _{i},\partial _{a})$ and
N--elongated differentials $\delta ^{\mu }=\delta u^{\mu }=(d^{i},\delta
^{a}).$ This emphasizes that the operators (\ref{dder}) and (\ref{ddif})
define certain ``N--elongated'' partial derivatives and differentials which
are more convenient for tensor and integral calculations on such
nonholonomic manifolds. The vielbeins (\ref{ddif}) satisfy the nonholo\-nomy
relations
\begin{equation}
\lbrack \mathbf{e}_{\alpha },\mathbf{e}_{\beta }]=\mathbf{e}_{\alpha }%
\mathbf{e}_{\beta }-\mathbf{e}_{\beta }\mathbf{e}_{\alpha }=W_{\alpha \beta
}^{\gamma }\mathbf{e}_{\gamma }  \label{anhrel}
\end{equation}%
with (antisymmetric) nontrivial anholonomy coefficients $W_{ia}^{b}=\partial
_{a}N_{i}^{b}$ and $W_{ji}^{a}=\Omega _{ij}^{a}$ defining a proper
parametrization (for a $n+m$ splitting by a N--connection $N_{i}^{a})$ of (%
\ref{ddif}).

\subsubsection{N--anholonomic manifolds and d--metrics}

For simplicity, we shall work with a particular class of nonholonomic
manifolds: A manifold $\mathbf{V}$ is \textbf{N--anholonomic} if its tangent
space $T\mathbf{V}$ is enabled with a N--connection structure (\ref{whitney}%
).\footnote{%
In a similar manner, we can consider different types of (super) spaces and
low energy string limits \cite{1vncsup,1vmon1,1vcv,1vstrf}, Riemann or
Riemann--Cartan manifolds \cite{1vsgg}, noncommutative bundles, or
superbundles and gauge models \cite{1vggr,1vdgrg,1dvgrg,1vgonch,1vesnc},
Clifford--Dirac spinor bundles and algebroids \cite%
{1vclalg,1vfs,1vhs,1vstav,1vv,1vtnut}, Lagrange--Fedosov manifolds \cite%
{1esv}... provided with nonholonomc (super) distributions (\ref{whitney})
and preferred systems of reference (supervielbeins).}

A \textbf{distinguished metric} (in brief, \textbf{d--metric}) on a
N--anholo\-nom\-ic manifold $\mathbf{V}$ is a usual second rank metric
tensor $\mathbf{g}$ which with respect to a N--adapted basis (\ref{ddif})
can be written in the form%
\begin{equation}
\mathbf{g}=\ g_{ij}(x,y)\ e^{i}\otimes e^{j}+\ h_{ab}(x,y)\ \mathbf{e}%
^{a}\otimes \mathbf{e}^{b}  \label{m1}
\end{equation}%
defining a N--adapted decomposition $\mathbf{g=}hg\mathbf{\oplus _{N}}%
vg=[hg,vg].$

A \textbf{metric structure } $\ \breve{g}$ on a N--anholonomic manifold $%
\mathbf{V}$ is a symmetric covariant second rank tensor field which is not
degenerated and of constant signature in any point $\mathbf{u\in V.}$ Any
metric on $\mathbf{V,}$ with respect to a local coordinate basis $du^{\alpha
}=\left( dx^{i},dy^{a}\right) ,$ can be parametrized in the form
\begin{equation}
\ \breve{g}=\underline{g}_{\alpha \beta }\left( u\right) du^{\alpha }\otimes
du^{\beta }  \label{metr}
\end{equation}%
where%
\begin{equation}
\underline{g}_{\alpha \beta }=\left[
\begin{array}{cc}
g_{ij}+N_{i}^{a}N_{j}^{b}h_{ab} & N_{j}^{e}h_{ae} \\
N_{i}^{e}h_{be} & h_{ab}%
\end{array}%
\right] .  \label{ansatz}
\end{equation}%
Such a metric (\ref{ansatz})\ is generic off--diagonal, i.e. it can not be
diagonalized by coordinate transforms if $N_{i}^{a}(u)$ are any general
functions.

In general, a metric structure is not adapted to a N--connection structure,
but we can transform it into a d--metric
\begin{equation}
\mathbf{g}=hg(hX,hY)+vg(vX,vY)  \label{dmetra}
\end{equation}%
adapted to a N--connection structure defined by coefficients $N_{i}^{a}.$ We
introduce denotations $h\breve{g}(hX,hY)=hg(hX,hY)$ and $v\breve{g}(vX,$ $%
vY)=vg(vX,vY)$ and try to find a N--connection when
\begin{equation}
\breve{g}(hX,vY)=0  \label{algn01}
\end{equation}%
for any d--vectors $\mathbf{X,Y.}$ In local form, for $hX\rightarrow e_{i}$
and $\ vY\rightarrow e_{a},$\ the equation (\ref{algn01}) is an algebraic
equation for the N--connection coefficients $N_{i}^{a},$
\begin{equation}
\breve{g}(e_{i},e_{a})=0, \mbox{ equivalently, } \underline{g}%
_{ia}-N_{i}^{b}h_{ab}=0,  \label{aux1a}
\end{equation}%
where $\underline{g}_{ia}$ $\doteqdot g(\partial /\partial x^{i},\partial
/\partial y^{a}),$ which allows us to define in a unique form the
coefficients $N_{i}^{b}=h^{ab}\underline{g}_{ia}$ where $h^{ab}$ is inverse
to $h_{ab}.$ We can write the metric $\breve{g}$ with ansatz (\ref{ansatz})\
in equivalent form, as a d--metric (\ref{m1}) adapted to a N--connection
structure, if we define $g_{ij}\doteqdot \mathbf{g}\left( e_{i},e_{j}\right)
$ and $h_{ab}\doteqdot \mathbf{g}\left( e_{a},e_{b}\right) $ \ and consider
the vielbeins $\mathbf{e}_{\alpha }$ and $\mathbf{e}^{\alpha }$ to be
respectively of type (\ref{dder}) and (\ref{ddif}).

A metric $\ \breve{g}$ (\ref{metr}) can be equivalently transformed into a
d--metric (\ref{m1}) by performing a frame (vielbein) transform
\begin{equation}
\mathbf{e}_{\alpha }=\mathbf{e}_{\alpha }^{\ \underline{\alpha }}\partial _{%
\underline{\alpha }}\mbox{ and }\mathbf{e}_{\ }^{\beta }=\mathbf{e}_{\
\underline{\beta }}^{\beta }du^{\underline{\beta }},  \label{ftnc}
\end{equation}%
with coefficients
\begin{eqnarray}
\mathbf{e}_{\alpha }^{\ \underline{\alpha }}(u) &=&\left[
\begin{array}{cc}
e_{i}^{\ \underline{i}}(u) & N_{i}^{b}(u)e_{b}^{\ \underline{a}}(u) \\
0 & e_{a}^{\ \underline{a}}(u)%
\end{array}%
\right] ,  \label{vt1} \\
\mathbf{e}_{\ \underline{\beta }}^{\beta }(u) &=&\left[
\begin{array}{cc}
e_{\ \underline{i}}^{i\ }(u) & -N_{k}^{b}(u)e_{\ \underline{i}}^{k\ }(u) \\
0 & e_{\ \underline{a}}^{a\ }(u)%
\end{array}%
\right] ,  \label{vt2}
\end{eqnarray}%
being linear on $N_{i}^{a}.$

It should be noted here that parametrizations of metrics of type (\ref%
{ansatz}) have been introduced in Kaluza--Klein gravity \cite{1ow} for the
case of linear connections (\ref{lincon}) and compactified extra dimensions $%
y^{a}.$ For the five (or higher) dimensions, the coefficients $\Gamma _{\
bi}^{a}(x)$ were considered as Abelian or non--Abelian gauge fields. In our
approach, the coefficients $N_{i}^{b}(x,y)$ are general ones, not obligatory
linearized and/or compactified on $y^{a}.$ For some models of Finsler
gravity, the values $N_{i}^{a}$ were treated as certain generalized
nonlinear gauge fields (see Appendix to Ref. \cite{1ma1987}), or as certain
objects defining (semi) spray configurations in generalized Finsler and
Lagrange gravity \cite{1ma1987,1ma,1aim}.

On N--anholonomic manifolds, we can say that the coordinates $x^{i}$ are
holonomic and the coordinates $y^{a}$ are nonholonomic (on N--anholonomic
vector bundles, such coordinates are called respectively to be the
horizontal and vertical ones). We conclude that a N--anholonomic manifold $%
\mathbf{V}$ provided with a metric structure $\breve{g}$ (\ref{metr})
(equivalently, with a d--metric (\ref{m1})) is a usual manifold (in
particular, a pseudo--Riemannian one) with a prescribed nonholonomic $n+m$
splitting into conventional ``horizontal'' and ``vertical'' subspaces (\ref%
{whitney}) induced by the ``off--diagonal'' terms $N_{i}^{b}(u)$ and the
corresponding preferred nonholonomic frame structure (\ref{anhrel}).

\subsubsection{d--torsions and d--curvatures}

From the general class of linear connections which can be defined on a
manifold $V,$ and any its N--anholonomic versions $\mathbf{V},$ we
distinguish those which are adapted to a N--connection structure $\mathbf{N.}
$

A \textbf{distinguished connection (d--connection) }$\mathbf{D}$ on a
N--anho\-lo\-no\-mic manifold $\mathbf{V}$ is a linear connection conserving
under parallelism the Whitney sum (\ref{whitney}). For any d--vector $%
\mathbf{X,}$ there is a decomposition of $\mathbf{D}$ into h-- and
v--covariant derivatives,%
\begin{equation}
\mathbf{D}_{\mathbf{X}}\mathbf{\doteqdot X}\rfloor \mathbf{D=}\ hX\rfloor
\mathbf{D+}\ vX\rfloor \mathbf{D=}Dh_{X}+D_{vX}=hD_{X}+vD_{X}.
\label{dconcov}
\end{equation}%
The symbol ''$\rfloor "$ in (\ref{dconcov}) denotes the interior product
defined by a metric (\ref{metr}) (equivalently, by a d--metric (\ref{m1})).
The N--adapted components $\mathbf{\Gamma }_{\ \beta \gamma }^{\alpha }$ of
a d--connection $\mathbf{D}_{\alpha }=(\mathbf{e}_{\alpha }\rfloor \mathbf{D}%
)$ are defined by the equations
\begin{equation}
\mathbf{D}_{\alpha }\mathbf{e}_{\beta }=\mathbf{\Gamma }_{\ \alpha \beta
}^{\gamma }\mathbf{e}_{\gamma },\mbox{\ or \ }\mathbf{\Gamma }_{\ \alpha
\beta }^{\gamma }\left( u\right) =\left( \mathbf{D}_{\alpha }\mathbf{e}%
_{\beta }\right) \rfloor \mathbf{e}^{\gamma }.  \label{dcon1}
\end{equation}%
The N--adapted splitting into h-- and v--covariant derivatives is stated by
\begin{equation*}
h\mathbf{D}=\{\mathbf{D}_{k}=\left( L_{jk}^{i},L_{bk\;}^{a}\right) \},%
\mbox{
and }\ v\mathbf{D}=\{\mathbf{D}_{c}=\left( C_{jc}^{i},C_{bc}^{a}\right) \},
\end{equation*}%
where $L_{jk}^{i}=\left( \mathbf{D}_{k}\mathbf{e}_{j}\right) \rfloor e^{i},
L_{bk}^{a}=\left( \mathbf{D}_{k}e_{b}\right) \rfloor \mathbf{e}^{a},
C_{jc}^{i}=\left( \mathbf{D}_{c}\mathbf{e}_{j}\right) \rfloor e^{i},
C_{bc}^{a}=\left( \mathbf{D}_{c}e_{b}\right) \rfloor \mathbf{e}^{a}.$ The
components $\mathbf{\Gamma }_{\ \alpha \beta }^{\gamma }=\left(
L_{jk}^{i},L_{bk}^{a},C_{jc}^{i},C_{bc}^{a}\right) $ completely define a
d--connec\-ti\-on $\mathbf{D}$ on a N--anholonomic manifold $\mathbf{V}.$ We
shall write conventionally that $\mathbf{D=}(hD,\ vD),$ or $\mathbf{D}%
_{\alpha }=(D_{i},D_{a}),$ with $hD=(L_{jk}^{i},L_{bk}^{a})$ and $%
vD=(C_{jc}^{i},$ $C_{bc}^{a}),$ see (\ref{dcon1}).

The \textbf{torsion and curvature} of a d--connection $\mathbf{D=}(hD,\ vD),$
\textbf{d--torsions and d--curvatures},\textbf{\ }are defined similarly to
formulas (\ref{ators}) and (\ref{acurv}) with further h-- and
v--decompositions. \ The simplest way to perform computations with
d--connections is to use \textbf{N--adapted differential forms} like
\begin{equation}
\mathbf{\Gamma }_{\ \beta }^{\alpha }=\mathbf{\Gamma }_{\ \beta \gamma
}^{\alpha }\mathbf{e}^{\gamma }  \label{dconf}
\end{equation}%
with the coefficients defined with respect to (\ref{ddif}) and (\ref{dder}).
For instance, torsion can be computed in the form
\begin{equation}
\mathcal{T}^{\alpha }\doteqdot \mathbf{De}^{\alpha }=d\mathbf{e}^{\alpha
}+\Gamma _{\ \beta }^{\alpha }\wedge \mathbf{e}^{\beta }.  \label{tors}
\end{equation}%
Locally it is characterized by (N--adapted) d--torsion coefficients
\begin{eqnarray}
T_{\ jk}^{i} &=&L_{\ jk}^{i}-L_{\ kj}^{i},\ T_{\ ja}^{i}=-T_{\ aj}^{i}=C_{\
ja}^{i},\ T_{\ ji}^{a}=\Omega _{\ ji}^{a},\   \notag \\
T_{\ bi}^{a} &=&-T_{\ ib}^{a}=\frac{\partial N_{i}^{a}}{\partial y^{b}}-L_{\
bi}^{a},\ T_{\ bc}^{a}=C_{\ bc}^{a}-C_{\ cb}^{a}.  \label{dtors}
\end{eqnarray}%
By a straightforward d--form calculus, we can compute the N--adapted
components $\mathbf{R=\{\mathbf{R}_{\ \beta \gamma \delta }^{\alpha }\}}$ of
the curvature
\begin{equation}
\mathcal{R}_{~\beta }^{\alpha }\doteqdot \mathbf{D\Gamma }_{\ \beta
}^{\alpha }=d\mathbf{\Gamma }_{\ \beta }^{\alpha }-\mathbf{\Gamma }_{\ \beta
}^{\gamma }\wedge \mathbf{\Gamma }_{\ \gamma }^{\alpha }=\mathbf{R}_{\ \beta
\gamma \delta }^{\alpha }\mathbf{e}^{\gamma }\wedge \mathbf{e}^{\delta },
\label{curv}
\end{equation}%
of a d--connection $\mathbf{D},$
\begin{eqnarray}
&&R_{\ hjk}^{i} =e_{k}L_{\ hj}^{i}-e_{j}L_{\ hk}^{i}+L_{\ hj}^{m}L_{\
mk}^{i}-L_{\ hk}^{m}L_{\ mj}^{i}-C_{\ ha}^{i}\Omega _{\ kj}^{a},  \notag \\
&&R_{\ bjk}^{a}=e_{k}L_{\ bj}^{a}-e_{j}L_{\ bk}^{a}+L_{\ bj}^{c}L_{\
ck}^{a}-L_{\ bk}^{c}L_{\ cj}^{a}-C_{\ bc}^{a}\Omega _{\ kj}^{c},
\label{dcurv} \\
&&R_{\ jka}^{i} =e_{a}L_{\ jk}^{i}-D_{k}C_{\ ja}^{i}+C_{\ jb}^{i}T_{\
ka}^{b},   \notag \\
&&R_{\ bka}^{c} = e_{a}L_{\ bk}^{c}-D_{k}C_{\ ba}^{c}+C_{\
bd}^{c}T_{\ ka}^{c},  \notag \\
&&R_{\ jbc}^{i} =e_{c}C_{\ jb}^{i}-e_{b}C_{\ jc}^{i}+C_{\ jb}^{h}C_{\
hc}^{i}-C_{\ jc}^{h}C_{\ hb}^{i},  \notag \\
&&R_{\ bcd}^{a} =e_{d}C_{\ bc}^{a}-e_{c}C_{\ bd}^{a}+C_{\ bc}^{e}C_{\
ed}^{a}-C_{\ bd}^{e}C_{\ ec}^{a}.  \notag
\end{eqnarray}

Contracting respectively the components of (\ref{dcurv}), one proves that
the Ricci tensor $\mathbf{R}_{\alpha \beta }\doteqdot \mathbf{R}_{\ \alpha
\beta \tau }^{\tau }$ is characterized by h- v--components
\begin{equation}
R_{ij}\doteqdot R_{\ ijk}^{k},\ \ R_{ia}\doteqdot -R_{\ ika}^{k},\
R_{ai}\doteqdot R_{\ aib}^{b},\ R_{ab}\doteqdot R_{\ abc}^{c}.
\label{dricci}
\end{equation}%
It should be noted that this tensor is not symmetric for arbitrary
d--connecti\-ons $\mathbf{D,}$ i.e. $\mathbf{R}_{\alpha \beta }\neq \mathbf{R%
}_{\beta \alpha }.$   The \textbf{scalar curvature} of a d--connection is
\begin{equation}
\ ^{s}\mathbf{R}\doteqdot \mathbf{g}^{\alpha \beta }\mathbf{R}_{\alpha \beta
}=g^{ij}R_{ij}+h^{ab}R_{ab},  \label{sdccurv}
\end{equation}%
defined by a sum the h-- and v--components of (\ref{dricci}) and d--metric (%
\ref{m1}).

The Einstein d--tensor is defined and computed similarly to (\ref{einstgr}),
but for d--connections,
\begin{equation}
\mathbf{E}_{\alpha \beta }=\mathbf{R}_{\alpha \beta }-\frac{1}{2}\mathbf{g}%
_{\alpha \beta }\ ^{s}\mathbf{R}  \label{enstdt}
\end{equation}%
This \ d--tensor defines an alternative to $\ _{\shortmid }E_{\alpha \beta }$
(nonholonomic) Einstein configuration if its d--connection is defined in a
unique form for an off--diagonal metric (\ref{ansatz}).

\subsubsection{Some classes of distinguished or non--adapted linear
connections}

From the class of arbitrary d--connections $\mathbf{D}$ on $\mathbf{V,}$ one
distinguishes those which are \textbf{metric compatible (metrical
d--connections)} satisfying the condition%
\begin{equation}
\mathbf{Dg=0}  \label{metcomp}
\end{equation}%
including all h- and v-projections $%
D_{j}g_{kl}=0,D_{a}g_{kl}=0,D_{j}h_{ab}=0,D_{a}h_{bc}=0.$ Different
approaches to Finsler--Lagrange geometry modelled on $\mathbf{TM}$ (or on
the dual tangent bundle $\mathbf{T}^{\ast }\mathbf{M,}$ in the case of
Cartan--Hamilton geometry) were elaborated for different d--metric
structures which are metric compatible \cite%
{1cart,1ma1987,1ma,1mhf,1mhss,1mhh,1vncsup,1vmon1,1vstav} or not metric
compatible \cite{1bcs}.

For any d--metric $\mathbf{g}=[hg,vg]$ on a N--anholonomic manifold $\mathbf{%
V,}$ there is a unique metric canonical d--connection $\widehat{\mathbf{D}}$
satisfying the conditions $\widehat{\mathbf{D}}\mathbf{g=}0$ and with
vanishing $h(hh)$--torsion, $v(vv)$--torsion, i. e. $h\widehat{T}(hX,hY)=0$
and $\mathbf{\ }v\widehat{T}(vX,\mathbf{\ }vY)=0.$ By straightforward
calculations, we can verify that $\widehat{\mathbf{\Gamma }}_{\ \alpha \beta
}^{\gamma }=\left( \widehat{L}_{jk}^{i},\widehat{L}_{bk}^{a},\widehat{C}%
_{jc}^{i},\widehat{C}_{bc}^{a}\right) ,$ when
\begin{eqnarray}
\widehat{L}_{jk}^{i} &=&\frac{1}{2}g^{ir}\left(
e_{k}g_{jr}+e_{j}g_{kr}-e_{r}g_{jk}\right) ,  \label{candcon} \\
\widehat{L}_{bk}^{a} &=&e_{b}(N_{k}^{a})+\frac{1}{2}h^{ac}\left(
e_{k}h_{bc}-h_{dc}\ e_{b}N_{k}^{d}-h_{db}\ e_{c}N_{k}^{d}\right) ,  \notag \\
\widehat{C}_{jc}^{i} &=&\frac{1}{2}g^{ik}e_{c}g_{jk},\ \widehat{C}_{bc}^{a}=%
\frac{1}{2}h^{ad}\left( e_{c}h_{bd}+e_{c}h_{cd}-e_{d}h_{bc}\right)  \notag
\end{eqnarray}%
result in $\widehat{T}_{\ jk}^{i}=0$ and $\widehat{T}_{\ bc}^{a}=0$ but $%
\widehat{T}_{\ ja}^{i},\widehat{T}_{\ ji}^{a}$ and $\widehat{T}_{\ bi}^{a}$
are not zero, see formulas (\ref{dtors}) written for this canonical
d--connection.

For any metric structure $\mathbf{g}$ on a manifold $\mathbf{V,}$ there is a
unique metric compatible and torsionless \textbf{Levi Civita connection} $%
\bigtriangledown =\{\ _{\shortmid }\Gamma _{\beta \gamma }^{\alpha }\}$ for
which $~\ _{\shortmid }\mathcal{T}=0$ and $\bigtriangledown g=0\mathbf{.}$
This is not a d--connection because it does not preserve under parallelism
the N--connection splitting (\ref{whitney}) (it is not adapted to the
N--connection structure). \ Let us parametrize its coefficients in the form
\begin{equation*}
_{\shortmid }\Gamma _{\beta \gamma }^{\alpha }=\left( _{\shortmid
}L_{jk}^{i},_{\shortmid }L_{jk}^{a},_{\shortmid }L_{bk}^{i},\ _{\shortmid
}L_{bk}^{a},_{\shortmid }C_{jb}^{i},_{\shortmid }C_{jb}^{a},_{\shortmid
}C_{bc}^{i},_{\shortmid }C_{bc}^{a}\right) ,
\end{equation*}%
where $\bigtriangledown _{\mathbf{e}_{k}}(\mathbf{e}_{j}) =\ _{\shortmid
}L_{jk}^{i}\mathbf{e}_{i}+\ _{\shortmid }L_{jk}^{a}e_{a},\ \bigtriangledown
_{\mathbf{e}_{k}}(e_{b})=\ _{\shortmid }L_{bk}^{i}\mathbf{e}_{i}+\
_{\shortmid }L_{bk}^{a}e_{a}, \bigtriangledown _{e_{b}}(\mathbf{e}_{j}) =\
_{\shortmid }C_{jb}^{i}\mathbf{e}_{i}+\ _{\shortmid }C_{jb}^{a}e_{a},\
\bigtriangledown _{e_{c}}(e_{b})=\ _{\shortmid }C_{bc}^{i}\mathbf{e}_{i}+\
_{\shortmid }C_{bc}^{a}e_{a}.$ A straightforward calculus\footnote{%
Such results were originally considered by R. Miron and M. Anastasiei for
vector bundles provided with N--connection and metric structures, see Ref. %
\cite{1ma}. Similar proofs hold true for any nonholonomic manifold provided
with a prescribed N--connection structure \cite{1vsgg}.} shows that the
coefficients of the Levi--Civita connection are%
\begin{eqnarray}
\ _{\shortmid }L_{jk}^{i} &=&L_{jk}^{i},\ _{\shortmid
}L_{jk}^{a}=-C_{jb}^{i}g_{ik}h^{ab}-\frac{1}{2}\Omega _{jk}^{a},  \notag \\
\ _{\shortmid }L_{bk}^{i} &=& \frac{1}{2}\Omega _{jk}^{c}h_{cb}g^{ji}-\frac{1%
}{2}(\delta _{j}^{i}\delta _{k}^{h}-g_{jk}g^{ih})C_{hb}^{j},  \notag \\
\ _{\shortmid }L_{bk}^{a} &=&L_{bk}^{a}+\frac{1}{2}(\delta _{c}^{a}\delta
_{d}^{b}+h_{cd}h^{ab})\left[ L_{bk}^{c}-e_{b}(N_{k}^{c})\right] ,
\label{lccon} \\
\ _{\shortmid }C_{kb}^{i} &=&C_{kb}^{i}+\frac{1}{2}\Omega
_{jk}^{a}h_{cb}g^{ji}+\frac{1}{2}(\delta _{j}^{i}\delta
_{k}^{h}-g_{jk}g^{ih})C_{hb}^{j},  \notag \\
\ _{\shortmid }C_{jb}^{a} &=&-\frac{1}{2}(\delta _{c}^{a}\delta
_{b}^{d}-h_{cb}h^{ad})\left[ L_{dj}^{c}-e_{d}(N_{j}^{c})\right] ,\
_{\shortmid }C_{bc}^{a}=C_{bc}^{a},  \notag \\
\ _{\shortmid }C_{ab}^{i} &=&-\frac{g^{ij}}{2}\left\{ \left[
L_{aj}^{c}-e_{a}(N_{j}^{c})\right] h_{cb}+\left[ L_{bj}^{c}-e_{b}(N_{j}^{c})%
\right] h_{ca}\right\} ,  \notag
\end{eqnarray}%
where $\Omega _{jk}^{a}$ are computed as in formula (\ref{ncurv}). For
certain considerations, it is convenient to express
\begin{equation}
\ _{\shortmid }\Gamma _{\ \alpha \beta }^{\gamma }=\widehat{\mathbf{\Gamma }}%
_{\ \alpha \beta }^{\gamma }+\ _{\shortmid }Z_{\ \alpha \beta }^{\gamma }
\label{cdeft}
\end{equation}%
where the explicit components of \textbf{distorsion tensor} $\ _{\shortmid
}Z_{\ \alpha \beta }^{\gamma }$ can be defined by comparing the formulas (%
\ref{lccon}) and (\ref{candcon}). It should be emphasized that all
components of $\ _{\shortmid }\Gamma _{\ \alpha \beta }^{\gamma },\widehat{%
\mathbf{\Gamma }}_{\ \alpha \beta }^{\gamma }$ and$\ _{\shortmid }Z_{\
\alpha \beta }^{\gamma }$ are uniquely defined by the coefficients of
d--metric (\ref{m1}) and N--connection (\ref{coeffnc}), or equivalently by
the coefficients of the corresponding generic off--diagonal metric\ (\ref%
{ansatz}).

\subsection{On equivalent (non)holonomic formulations of gra\-vity theories}

A \textbf{N--anholonomic Riemann--Cartan manifold} $\ ^{RC}\mathbf{V}$ is
defined by a d--metric $\mathbf{g}$ and a metric d--connection $\mathbf{D}$
structures. We can say that a space$\ ^{R}\widehat{\mathbf{V}}$ is a
canonical N--anholonomic Riemann manifold if its d--connecti\-on structure
is canonical, i.e. $\mathbf{D=}\widehat{\mathbf{D}}.$ The d--metric
structure $\mathbf{g}$ on$\ ^{RC}\mathbf{V}$ is of type (\ref{m1}) and
satisfies the metricity conditions (\ref{metcomp}). With respect to a local
coordinate basis, the metric $\mathbf{g}$ is parametrized by a generic
off--diagonal metric ansatz (\ref{ansatz}). For a particular case, we can
treat the torsion $\widehat{\mathbf{T}}$ as a nonholonomic frame effect
induced by a nonintegrable N--splitting. We conclude that a manifold $^{R}%
\widehat{\mathbf{V}}$ is enabled with a nontrivial torsion (\ref{dtors})
(uniquely defined by the coefficients of N--connection (\ref{coeffnc}), and
d--metric (\ref{m1}) and canonical d--connection (\ref{candcon})
structures). Nevertheless, such manifolds can be described alternatively,
equivalently, as a usual (holonomic) Riemann manifold \ with the usual Levi
Civita for the metric (\ref{metr}) with coefficients (\ref{ansatz}). We do
not distinguish the existing nonholonomic structure for such geometric
constructions.

Having prescribed a nonholonomic $n+m$ splitting on a manifold $V,$ we can
define two canonical linear connections $\nabla $ and $\widehat{\mathbf{D}}.$
Correspondingly, these connections are characterized by two curvature
tensors, $_{\shortmid }R_{~\beta \gamma \delta }^{\alpha }(\nabla )$
(computed by introducing $_{\shortmid }\Gamma _{\beta \gamma }^{\alpha }$
into (\ref{dconf}) and (\ref{curv})) and $\mathbf{R}_{~\beta \gamma \delta
}^{\alpha }(\widehat{\mathbf{D}})$ (with the N--adapted coefficients
computed following formulas (\ref{dcurv})). Contracting indices, we can
commute the Ricci tensor $Ric(\nabla )$ and the Ricci d--tensor $\mathbf{Ric}%
(\widehat{\mathbf{D}})$ following formulas (\ref{dricci}), correspondingly
written for $\nabla $ and $\widehat{\mathbf{D}}.$ Finally, using the inverse
d--tensor $\mathbf{g}^{\alpha \beta }$ for both cases, we compute the
corresponding scalar curvatures $\ ^{s}R(\nabla )$ and $\ ^{s}\mathbf{R(%
\widehat{\mathbf{D}}),}$ see formulas (\ref{sdccurv}) by contracting,
respectively, with the Ricci tensor and Ricci d--tensor.

The standard formulation of the Einstein gravity is for the connection $%
\nabla ,$ when the field equations are written in the form (\ref{einstgr}).
But it can be equivalently reformulated by using the canonical
d--connection, or other connections uniquely defined by the metric
structure. If a metric (\ref{ansatz}) $\underline{g}_{\alpha \beta }$ is a
solution of the Einstein equations$\ _{\shortmid }E_{\alpha \beta }=\Upsilon
_{\alpha \beta },$ having prescribed a $\left( n+m\right) $--decomposition,
we can define algebraically the coefficients of a N--connection, $N_{i}^{a},$
N--adapted frames $e_{\alpha }$ (\ref{dder}) and $e^{\beta }$ (\ref{ddif}),
and d--metric $\mathbf{g}_{\alpha \beta }=[g_{ij},h_{ab}]$ (\ref{m1}). The
next steps are to compute $\widehat{\mathbf{\Gamma }}_{\ \alpha \beta
}^{\gamma }, $ following formulas (\ref{candcon}), and then using (\ref%
{dcurv}), (\ref{dricci}) and (\ref{sdccurv}) for $\widehat{\mathbf{D}},$ to
define $\widehat{\mathbf{E}}_{\alpha \beta }$ (\ref{enstdt}). The Einstein
equations with matter sources, written in equivalent form by using the
canonical d--connection, are
\begin{equation}
\widehat{\mathbf{E}}_{\alpha \beta }=\mathbf{\Upsilon }_{\alpha \beta }+\
^{Z}\mathbf{\Upsilon }_{\alpha \beta },  \label{einstgrdef}
\end{equation}%
where the effective source $\ ^{Z}\mathbf{\Upsilon }_{\alpha \beta }$ is
just the deformation tensor of the Einstein tensor computed by introducing
deformation (\ref{cdeft}) into the left part of (\ref{einstgr}); all
decompositions being performed with respect to the N--adapted co--frame (\ref%
{ddif}), when $_{\shortmid }E_{\alpha \beta }=\widehat{\mathbf{E}}_{\alpha
\beta }-\ ^{Z}\mathbf{\Upsilon }_{\alpha \beta }.$ For certain matter field/
string gravity configurations, the solutions of (\ref{einstgrdef}) also
solve the equations (\ref{einstgr}). Nevertheless, because of generic
nonlinear character of gravity and gravity--matter field interactions and
functions defining nonholonomic distributions, one could be certain special
conditions when even vacuum configurations contain a different physical
information if to compare with usual holonomic ones. We analyze some
examples:

In our works \cite{1vesnc,1vsgg,1vclalg}, we investigated a series of exact
solutions defining N--anholonomic Einstein spaces related to generic
off--diagonal solutions in general relativity by such nonholonomic
constraints when $\mathbf{Ric}(\widehat{\mathbf{D}})=Ric(\mathbf{\nabla }),$
even $\widehat{\mathbf{D}}\neq \nabla .$\footnote{%
One should be emphasized here that different type of connections on
N--anholonomic manifolds have different coordinate and frame transform
properties. It is possible, for instance, to get equalities of coefficients
for some systems of coordinates even the connections are very different. The
transformation laws of tensors and d--tensors are also different if some
objects are adapted and other are not adapted to a prescribed N--connection
structure.} In this case, for instance, the solutions of the Einstein
equations with cosmological constant $\lambda ,$%
\begin{equation}
\widehat{\mathbf{R}}_{\alpha \beta }=\lambda \mathbf{g}_{\alpha \beta }
\label{nhesp}
\end{equation}%
can be transformed into metrics for usual Einstein spaces with Levi Civita
connection $\nabla .$ The idea is that for certain general metric ansatz,
see section \ref{exsolans}, the equations (\ref{nhesp}) can be integrated in
general form just for the connection $\widehat{\mathbf{D}}$ but not for $%
\nabla .$ The nontrivial torsion components
\begin{equation}
\ \widehat{T}_{\ ja}^{i}==-\widehat{T}_{\ aj}^{i}=\widehat{C}_{\ ja}^{i},\
\widehat{T}_{\ ji}^{a}=\ \widehat{T}_{\ ij}^{a}=\Omega _{\ ji}^{a},\
\widehat{T}_{\ bi}^{a}=-\widehat{T}_{\ ib}^{a}=\frac{\partial N_{i}^{a}}{%
\partial y^{b}}-\widehat{L}_{\ bi}^{a},  \label{dtorsc}
\end{equation}%
see (\ref{dtors}), for some configurations, may be associated with an
absolute antisymmetric $H$--fields in string gravity \cite{1string1,1string2}%
, but nonholonomically transformed to N--adapted bases, see details in \cite%
{1vesnc,1vsgg}.

For more restricted configurations, we can search solutions with metric
ansatz defining Einstein foliated spaces, when
\begin{equation}
\Omega _{jk}^{c}=0,\ \widehat{L}_{bk}^{c}=e_{b}(N_{k}^{c}),\ \widehat{C}%
_{jb}^{i}=0,  \label{intcond}
\end{equation}%
and the d--torsion components (\ref{dtorsc}) vanish, but the N--adapted
frame structure has, in general, nontrivial anholonomy coefficients, see (%
\ref{anhrel}). One present a special interest a less constrained
configurations with $\ \widehat{T}_{\ jk}^{c}=\Omega _{jk}^{c}\neq 0$ when $%
\mathbf{Ric}(\widehat{\mathbf{D}})=Ric(\mathbf{\nabla })$ and $\widehat{T}%
_{jk}^{i}=\widehat{T}_{bc}^{a}=0,$ for certain general ansatz $\widehat{T}%
_{\ ja}^{i}=0$ and $\widehat{T}_{\ bi}^{a}=0,$ but $\widehat{\mathbf{\mathbf{%
R}}}\mathbf{_{\ \beta \gamma \delta }^{\alpha }\neq \ }_{\shortmid }R\mathbf{%
_{\ \beta \gamma \delta }^{\alpha }.}$ In such cases, we constrain the
integral varieties of equations (\ref{nhesp}) in such a manner that we
generate integrable or nonintegrable distributions on a usual Einstein space
defined by $\nabla .$ This is possible because if the conditions (\ref%
{intcond}) are satisfied, the deformation tensor $\ _{\shortmid }Z_{\ \alpha
\beta }^{\gamma }=0.$ For $\lambda =0,$ if $n+m=4,$ for corresponding
signature, we get foliated vacuum configurations in general relativity.

For N--anholonomic manifolds $\mathbf{V}^{n+n}$ of odd dimensions, when $%
m=n, $ and if $g_{ij}=h_{ij}$ (we identify correspondingly, the h- and
v--indices), we can consider a canonical d--connection$\ \widehat{\mathbf{D}}%
=(h\widehat{D},v\widehat{D})$ with the nontrivial coefficients with respect
to $\mathbf{e}_{\nu }$ and $\mathbf{e}^{\mu }$ paramet\-riz\-ed respectively
$\widehat{\mathbf{\Gamma }}_{\ \beta \gamma }^{\alpha }=(\widehat{L}_{\
jk}^{i}=\widehat{L}_{\ bk}^{a},\widehat{C}_{jc}^{i}=\widehat{C}_{bc}^{a}),$%
\footnote{%
the equalities of indices ''$i=a"$ are considered in the form $"i=1=a=n+1,$ $%
i=2=a=n+2,$ ... $i=n=a=n+n"$} for 
\begin{equation}
\widehat{L}_{\ jk}^{i} =\frac{1}{2}g^{ih}(\mathbf{e}_{k}g_{jh}+\mathbf{e}%
_{j}g_{kh}-\mathbf{e}_{h}g_{jk}), \widehat{C}_{\ bc}^{a} =\frac{1}{2}%
g^{ae}(e_{b}g_{ec}+e_{c}g_{eb}-e_{e}g_{bc}),  \label{3cdctb}
\end{equation}
defining the generalized Christoffel symbols. Such nonholonomic
configurations can be used for modelling generalized Finsler--Lagrange, and
particular cases, defined in Refs. \cite{1ma1987,1ma} for $\mathbf{V}^{n+n}=%
\mathbf{TM,}$ see below section \ref{sfls}. \ There are only three classes
of d--curvatures for the d--connection (\ref{3cdctb}),%
\begin{eqnarray}
\widehat{R}_{\ hjk}^{i} &=&\mathbf{e}_{k}\widehat{L}_{\ hj}^{i}-\mathbf{e}%
_{j}\widehat{L}_{\ hk}^{i}+\widehat{L}_{\ hj}^{m}\widehat{L}_{\ mk}^{i}-%
\widehat{L}_{\ hk}^{m}\widehat{L}_{\ mj}^{i}-\widehat{C}_{\ ha}^{i}\Omega
_{\ kj}^{a},  \label{3dcurvtb} \\
\widehat{P}_{\ jka}^{i} &=&e_{a}\widehat{L}_{\ jk}^{i}-\widehat{\mathbf{D}}%
_{k}\widehat{C}_{\ ja}^{i},\ \widehat{S}_{\ bcd}^{a}=e_{d}\widehat{C}_{\
bc}^{a}-e_{c}\widehat{C}_{\ bd}^{a}+\widehat{C}_{\ bc}^{e}\widehat{C}_{\
ed}^{a}-\widehat{C}_{\ bd}^{e}\widehat{C}_{\ ec}^{a},  \notag
\end{eqnarray}%
where all indices $a,b,...,i,j,...$ run the same values and, for instance, $%
C_{\ bc}^{e}\rightarrow $ $C_{\ jk}^{i},...$ Such locally anisotropic
configurations are not integrable if $\Omega _{\ kj}^{a}\neq 0,$ even the
d--torsion components $\widehat{T}_{\ jk}^{i}=0$ and $\widehat{T}_{\
bc}^{a}=0.$ We note that for geometric models on $\mathbf{V}^{n+n},$ or on $%
\mathbf{TM,}$ with $g_{ij}=h_{ij},$ one writes, in brief, $\widehat{\mathbf{%
\Gamma }}_{\ \beta \gamma }^{\alpha }=\left( \widehat{L}_{\ jk}^{i},\widehat{%
C}_{\ bc}^{a}\right) ,$ or, for more general d--connections, $\mathbf{\Gamma
}_{\ \beta \gamma }^{\alpha }=\left( L_{\ jk}^{i},C_{\ bc}^{a}\right) ,$ see
below section \ref{sfls}, on Lagrange and Finsler spaces.

\section{Nonholonomic Deformations of Manifolds and Vector Bundles}

This section will deal mostly with nonholonomic distributions on manifolds
and vector/ tangent bundles and their nonholonomic deformations modelling,
on Riemann and Riemann--Cartan manifolds, different types of generalized
Finsler--Lagrange geometries.

\subsection{Finsler--Lagrange spaces and generalizations}

\label{sfls}The notion of Lagrange space was introduced by J. Kern \cite%
{1kern} and elaborated in details by R. Miron's school, see Refs. \cite%
{1ma1987,1ma,1mhl,1mhf,1mhss,1mhh}, as a natural extension of Finsler
geometry \cite{1cart,1rund,1mats,1bej} (see also Refs. \cite%
{1vncsup,1vmon1,1vcv,1vesnc,1vsgg}, on Lagrange--Finsler
super/noncommutative geometry). Originally, such geometries were constructed
on tangent bundles, but they also can be modelled on N--anholonomic
manifolds, for instance, as models for certain gravitational interactions
with prescribed nonholonomic constraints deformed symmetries.

\subsubsection{Lagrange spaces}

A \textbf{differentiable Lagrangian} $L(x,y),$ i.e. a fundamental Lagrange
function, is defined by a map $L:(x,y)\in TM\rightarrow L(x,y)\in \mathbb{R}$
of class $\mathcal{C}^{\infty }$ on $\widetilde{TM}=TM\backslash \{0\}$ and
continuous on the null section $0:\ M\rightarrow TM$ of $\pi .$ A regular
Lagrangian has non-degenerate \textbf{Hessian}
\begin{equation}
\ ^{L}g_{ij}(x,y)=\frac{1}{2}\frac{\partial ^{2}L(x,y)}{\partial
y^{i}\partial y^{j}},  \label{lqf}
\end{equation}%
when $rank\left| g_{ij}\right| =n$ and $^{L}g^{ij}$ is the inverse matrix. A
\textbf{Lagrange space }is a pair $L^{n}=\left[ M,L(x,y)\right] $ with $\
^{L}g_{ij}$ being of fixed signature over $\mathbf{V}=\widetilde{TM}.$

One holds the result:\ The \textbf{Euler--Lagrange equations}%
 $\frac{d}{d\tau }\left( \frac{\partial L}{\partial y^{i}}\right) -\frac{%
\partial L}{\partial x^{i}}=0,$ where $y^{i}=\frac{dx^{i}}{d\tau }$ for $x^{i}(\tau )$ depending on
parameter $\tau ,$ are equivalent to the \textbf{``nonlinear'' geodesic
equations} 
$\frac{d^{2}x^{a}}{d\tau ^{2}}+2G^{a}(x^{k},\frac{dx^{b}}{d\tau })=0$
defining paths of a canonical \textbf{semispray}
 $S=y^{i}\frac{\partial }{\partial x^{i}}-2G^{a}(x,y)\frac{\partial }{\partial
y^{a}},$  where $2G^{i}(x,y)=\frac{1}{2}\ ^{L}g^{ij}\left( \frac{%
\partial ^{2}L}{\partial y^{i}\partial x^{k}}y^{k}-\frac{\partial L}{%
\partial x^{i}}\right).$

There exists on $\mathbf{V\simeq }$ $\widetilde{TM}$ a canonical
N--connection $\ $%
\begin{equation}
\ ^{L}N_{j}^{a}=\frac{\partial G^{a}(x,y)}{\partial y^{j}}  \label{cncl}
\end{equation}%
defined by the fundamental Lagrange function $L(x,y),$ which prescribes
nonholonomic frame structures of type (\ref{dder}) and (\ref{ddif}), $^{L}%
\mathbf{e}_{\nu }=(\ ^{L}\mathbf{e}_{i},e_{a})$ and $^{L}\mathbf{e}^{\mu
}=(e^{i},\ ^{L}\mathbf{e}^{a}).$ One defines the canonical metric structure%
\begin{equation}
\ ^{L}\mathbf{g}=\ ^{L}g_{ij}(x,y)\ e^{i}\otimes e^{j}+\ ^{L}g_{ij}(x,y)\
^{L}\mathbf{e}^{i}\otimes \ ^{L}\mathbf{e}^{j}  \label{slm}
\end{equation}%
constructed as a Sasaki type lift from $M$ for $\ ^{L}g_{ij}(x,y),$ see
details in \cite{1yano,1ma1987,1ma}.

There is a unique canonical d--connection$\ ^{L}\widehat{\mathbf{D}}=(h\ ^{L}%
\widehat{D},v\ ^{L}\widehat{D})$ with the coefficients $\ ^{L}\widehat{%
\Gamma }_{\ \beta \gamma }^{\alpha }=(\ ^{L}\widehat{L}_{\ jk}^{i},\ ^{L}%
\widehat{C}_{bc}^{a})$ computed by formulas (\ref{3cdctb}) for the d--metric
(\ref{slm}) with respect to $^{L}\mathbf{e}_{\nu }$ and $^{L}\mathbf{e}^{\mu
}.$ All such geometric objects, including the corresponding to $^{L}\widehat{%
\Gamma }_{\ \beta \gamma }^{\alpha },\ ^{L}\mathbf{g}$ and $^{L}N_{j}^{a}$
d--curvatures\newline
$\ ^{L}\widehat{\mathbf{\mathbf{R}}}\mathbf{_{\ \beta \gamma \delta
}^{\alpha }=}\left( \ ^{L}\widehat{R}_{\ hjk}^{i},\ ^{L}\widehat{P}_{\
jka}^{i},\ ^{L}\widehat{S}_{\ bcd}^{a}\right) ,$ see (\ref{3dcurvtb}), are
completely defined by a Lagrange fundamental function $L(x,y)$ for a
nondegerate $^{L}g_{ij}.$

We conclude that any regular Lagrange mechanics can be geometrized as a
nonholonomic Riemann manifold $^{L}\mathbf{V}$ equipped with the canonical
N--connecti\-on $^{L}N_{j}^{a}$ (\ref{cncl}). This geometrization was
performed in such a way that the N--connecti\-on is induced canonically by
the semispray configurations subjected the condition that the generalized
nonlinear geodesic equations are equivalent to the Euler--Lagrange equations
for $L.$ Such mechanical models and semispray configurations can be used for
a study of certain classes of nonholonomic effective analogous of
gravitational interactions. The approach can be extended for more general
classes of effective metrics, then those parametrized by (\ref{slm}), see
next sections. After Kern and Miron and Anastasiei works, it was elaborated
the so--called ''analogous gravity'' approach \cite{1blv} with similar ideas
modelling related to continuous mechanics, condensed media.... It should be
noted here, that the constructions for higher order generalized Lagrange and
Hamilton spaces \cite{1mhl,1mhf,1mhss,1mhh} provided a comprehensive
geometric formalism for analogous models in gravity, geometric mechanics,
continuous media, nonhomogeneous optics etc etc.

\subsubsection{Finsler spaces}

Following the ideas of the Romanian school on Finsler--Lagrange geometry and
generalizations, any Finsler space defined by a \textbf{fundamental Finsler
function} $F(x,y),$ being homogeneous of type $F(x,\lambda y)=|\lambda |\
F(x,y),$ for nonzero $\lambda \in \mathbb{R},$ may be considered as a
particular case of Lagrange space when $L=F^{2}$ (on different rigorous
mathematical definitions of Finsler spaces, see \cite%
{1rund,1mats,1ma1987,1ma,1bej,1bcs}; in our approach with applications to
physics, we shall not constrain ourself with special signatures, smooth
class conditions and special types of connections). Historically, the bulk
of mathematicians worked in an inverse direction, by generalizing the
constructions from the Cartan's approach to Finsler geometry in order to
include into consideration regular Lagrange mechanical systems, or to define
Finsler geometries with another type of nonlinear and linear connection
structures. The Finsler geometry, in terms of the normal canonical
d--connection (\ref{3cdctb}), derived for respective $\ ^{F}g_{ij}$ and $%
^{F}N_{j}^{a},$ can be modelled as for the case of Lagrange spaces
consudered in the previous section: we have to change formally all labels $%
L\rightarrow F$ $\ $and take into consideration possible conditions of
homogeneity (or $TM,$ see the monographs \cite{1ma1987,1ma}).

For generalized Finsler spaces, a N--connection can be stated by a general
set of coeficients $N_{j}^{a}$ subjected to certain nonholonomy conditions.
Of course, working with homogeneous functions on a manifold $V^{n+n},$ we
can model a Finsler geometry both on holonomic and nonholonomic Riemannian
manifolds, or on certain types of Riemann--Cartan manifolds enabled with
preferred frame structures $^{F}\mathbf{e}_{\nu }=(\ ^{F}\mathbf{e}%
_{i},e_{a})$ and $^{F}\mathbf{e}^{\mu }=(e^{i},\ ^{F}\mathbf{e}^{a}).$
Bellow, in the section \ref{exsolans}, we shall discuss how certain type
Finsler configurations can be derived as exact solutions in Einstein
gravity. Such constructions allow us to argue that Finsler geometry is also
very important in standard physics and that it was a big confusion to treat
it only as a ''sophisticated'' generalization of Riemann geometry, on
tangent bundles, with not much perspectives for modern physics.

In a number of works (see monographs \cite{1ma1987,1ma,1bcs}), it is
emphasized that the first example of Finsler metric was considered in the
famous inauguration thesis of B. Riemann \cite{1riemann}, long time before
P.\ Finsler \cite{1fg}. Perhaps, this is a reason, for many authors, to use
the term Riemann--Finsler geometry. Nevertheless, we would like to emphasize
that a Finsler space is not completely defined only by a metric structure of
type
\begin{equation}
\ ^{F}g_{ij}=\frac{1}{2}\frac{\partial ^{2}F}{\partial y^{i}\partial y^{j}}
\label{fhes}
\end{equation}%
originally considered on the vertical fibers of a tangent bundle. There are
necessary additional conventions about metrics on a total Finsler space,
N--connections and linear connections. This is the source for different
approaches, definitions, constructions and ambiguities related to Finsler
spaces and applications. Roughly speaking, different famous mathematicians,
and their schools, elaborated their versions of Finsler geometries following
some special purposes in geometry, mechanics and physics.

The first complete model of Finsler geometry exists due to E. Cartan \cite%
{1cart} who in the 20-30th years of previous century elaborated the concepts
of vector bundles, Rieman--Cartan spaces with torsion, moving frames,
developed the theory of spinors, Pfaff forms ... and (in coordinate form)
operated with nonlinear connection coefficients. The Cartan's constructions
were performed with metric compatible linear connections which is very
important for applications to standard models in physics.

Latter, there were proposed different models of Finsler spaces with metric
not compatible linear connections. The most notable connections were those
by L. Berwald, S. -S. Chern (re--discovered by H. Rund), H. Shimada and
others (see details, discussions and bibliography in monographs \cite%
{1ma1987,1ma,1bcs,1rund}). For d--connections of type (\ref{3cdctb}), there
are distinguished three cases of metric compatibility (compare with h- and
v-projections of formula (\ref{metcomp})):\ A Finsler connection $^{F}%
\mathbf{D}_{\alpha }=(\ ^{F}D_{k},\ ^{F}D_{a})$ is called h--metric if $%
^{F}D_{i}^{F}g_{ij}=0;$ it is called v--metric if $^{F}D_{a}\ ^{F}g_{ij}=0$
and it is metrical if both conditions are satisfied.

Here, we note three of the most important Finsler d--connections having their
special geometric and (possible) physical merits:
\begin{enumerate}
\item The \textbf{canonical Finsler connection} $\ ^{F}\widehat{\mathbf{D}}$
is defined by formulas (\ref{3cdctb}), but for $^{F}g_{ij},$ i.e. as $\ ^{F}%
\widehat{\mathbf{\Gamma }}_{\ \beta \gamma }^{\alpha }=\left( \ ^{F}\widehat{%
L}_{\ jk}^{i},\ ^{F}\widehat{C}_{\ bc}^{a}\right) .$ This d--connection is
metrical. For a special class of N--connections $%
^{C}N_{j}^{a}(x^{k},y^{b})=y^{k}\ ^{C}L_{\ kj}^{i},$ we get the famous
\textbf{Cartan connection} for Finsler spaces, $\ ^{C}\mathbf{\Gamma }_{\
\beta \gamma }^{\alpha }=\left( \ ^{C}L_{\ jk}^{i},\ ^{C}C_{\ bc}^{a}\right)
,$ with
\begin{eqnarray}
\ ^{C}L_{\ jk}^{i} &=&\frac{1}{2}\ ^{F}g^{ih}(\ ^{C}\mathbf{e}_{k}\
^{F}g_{jh}+\ ^{C}\mathbf{e}_{j}\ ^{F}g_{kh}-\ ^{C}\mathbf{e}_{h}\
^{F}g_{jk}),  \label{cartan} \\
\ ^{C}C_{\ bc}^{a} &=&\frac{1}{2}\ ^{F}g^{ae}(e_{b}\ ^{F}g_{ec}+e_{c}\
^{F}g_{eb}-e_{e}\ ^{F}g_{bc}),  \notag
\end{eqnarray}%
where $\ ^{C}\mathbf{e}_{k}=\frac{\partial }{\partial x^{k}}-\ ^{C}N_{j}^{a}%
\frac{\partial }{\partial y^{a}}$ and $e_{b}=\frac{\partial }{\partial y^{b}}%
,$ which can be defined in a unique axiomatic form \cite{1mats}. Such
canonical and Cartan--Finsler connections, being metric compatible, for
nonholonomic geometric models with local anisotropy on Riemann or
Riemann--Cartan manifolds, are more suitable with the paradigm of modern
standard physics.

\item The \textbf{Berwald connection} $\ ^{B}\mathbf{D}$ was introduced in
the form $\ ^{B}\mathbf{\Gamma }_{\ \beta \gamma }^{\alpha }=\left( \frac{%
\partial \ ^{C}N_{j}^{b}}{\partial y^{a}},0\right)$ \cite{1bw}. This
d--connection is defined completely by the N--connec\-tion structure but it
is not metric compatible, both not h--metric and not v--metric.

\item The \textbf{Chern connection} $^{Ch}\mathbf{D}$ was considered as a
minimal Finsler extension of the Levi Civita connection, $\ ^{Ch}\mathbf{%
\Gamma }_{\ \beta \gamma }^{\alpha }=\left( \ ^{C}L_{\ jk}^{i},0\right) ,$
with $\ ^{C}L_{\ jk}^{i}$ defined as in (\ref{cartan}), preserving the
torsionless condition, being h--metric but not v--metric. It is an
interesting case of nonholonomic geometries when torsion is completely
transformed into nonmetricity which for physicists presented a substantial
interest in connection to the Weyl nonmetricity introduced as a method of
preserving conformal symmetry of certain scalar field constructions in
general relativity, see discussion in \cite{1mag}. Nevertheless, it should
be noted that the constructions with the Chern connection, in general, are
not metric compatible and can not be applied in direct form to standard
models of physics.
\end{enumerate}

It should be noted that all mentioned types of d--connections are uniquely
defined by the coefficients of Finsler type d--metric and N--connection
structure (equivalently, by the coefficients of corresponding generic
off--diagonal metric of type (\ref{ansatz})) following well defined
geometric conditions. From such d--connections, we can always 'extract' the
Levi Civita connection, using formulas of type (\ref{lccon}) and (\ref{cdeft}%
), and work in 'non--adapted' (to N--connection) form. From geometric point
of view, we can work with all types of Finsler connections and elaborate
equivalent approaches even different connections have different merits in
some directions of physics. For instance, in \cite{1ma1987,1ma}, there are
considered the Kawaguchi metrization procedure and the Miron's method of
computing all metric compatible Finsler connections starting with a
canonical one. It was analyzed also the problem of transforming one Finsler
connection into different ones on tangent bundles and the formalism of
mutual transforms of connections was reconsidered for nonholonomic
manifolds, see details in \cite{1vsgg}.

Different models of Finsler spaces can be elaborated in explicit form for
different types of d--metrics, N--connections and d--connections. For
instance, for a Finsler Hessian (\ref{fhes}) defining a particular case of
d--metrics (\ref{slm}), or (\ref{m1}), denoted $\ ^{F}\mathbf{g,}$ for any
type of connection (for instance, canonical d--connection, Cartan--Finsler,
Berwald, Chern etc), we can compute the curvatures by using
formulas (\ref{3dcurvtb}) when ''hat'' labels are changed into the
corresponding ones ''$C,B,Ch,...".$ This way, we model Finsler geometries
on tangent bundles, like it is considered in the bulk of monographs \cite%
{1cart,1mats,1rund,1ma1987,1ma,1bej,1bcs}, or on nonholonomic manifolds \cite%
{1vr1,1vr1a,1vr2,1hor,1bejf,1vsgg}.

With the aim to develop new applications in standard models of physics, let
say in classical general relativity, when Finsler like structures are
modelled on a (pseudo) Riemannian manifold (we shall consider explicit
examples in the next sections), it is positively sure that the canonical
Finsler and Cartan connections, and their variants of canonical
d--connection on vector bundles and nonholonomic manifolds, should be
preferred for constructing new classes of Einstein spaces and defining
certain low energy limits to locally anisotropic string gravity models. Here
we note that it is a very difficult problem to define Finsler--Clifford
spaces with Finsler spinors, noncommutative generalizations to
supersymmetric/ noncommutative Finsler geometry if we work with nonmetric
d--connections, see discussions in \cite{1vsgg,1vstav,1vmon1}.

We cite a proof \cite{1bcs} that any Lagrange fundamental function $L$ can
be modelled as a singular case for a certain class of Finsler geometries of
extra dimension (perhaps, the authors were oriented to prove from a
mathematical point of view that it is not necessary to develop Finsler
geometry as a new theory for Lagrange spaces, or their dual constructions
for the Hamilton spaces). This idea, together with the method of
Kawaguchi--Miron transforms of connections, can be related to the H.
Poincare philosophical concepts about conventionality of the geometric space
and field interaction theories \cite{1poinc1,1poinc2}. According to the
Poincare's geometry--physics dualism, the procedure of choosing a geometric
arena for a physical theory is a question of convenience for researches to
classify scientific data in an economical way, but not an action to be
verified in physical experiments: as a matter of principe, any physical
theory can be equivalently described on various types of geometric spaces by
using more or less "simple" geometric objects and transforms.

Nevertheless, the modern physics paradigm is based on the ideas of objective
reality of physical laws and their experimental and theoretical
verifications, at least in indirect form. The concept of Lagrangian is a
very important geometrical and physical one and we shall distinguish the
cases when we model a Lagrange or a Finsler geometry. A physical or
mechanical model with a Lagrangian is not only a ''singular'' case for a
Finsler geometry but reflects a proper set of \ concepts, fundamental
physical laws and symmetries describing real physical effects. We use the
terms Finsler and Lagrange spaces in order to emphasize that they are
different both from geometric and physical points of view. Certain geometric
concepts and methods (like the N--connection geometry and nonholonomic frame
transforms ...) are very important for both types of geometries, modelled on
tangent bundles or on nonholonomic manifolds. This will be noted when we use
the term Finsler--Lagrange geometry (structures, configurations, spaces).

One should be emphasized that the author of this review should not be
considered as a physicist who does not accept nonmetric geometric
constructions in modern physics. For instance, the Part I in monograph \cite%
{1vsgg} is devoted to a deep study of the problem when generalized
Finsler--Lagrange structures can be modelled on metric--affine spaces, even
as exact solutions in gravity with nonmetricity \cite{1mag}, and, inversely,
the Lagrange--affine and Finsler--affine spaces are classified by
nonholonomic structures on metric--affine spaces. It is a question of
convention on the type of physical theories one models by geometric methods.
The standard theories of physics are formulated for metric compatible
geometries, but further developments in quantum gravity may request certain
type of nonmetric Finsler like geometries, or more general constructions.
This is a topic for further investigations.

\subsubsection{Generalized Lagrange spaces}

There are various application in optics of nonhomogeneous media and gravity
(see, for instance, Refs. \cite{1ma,1vsgg,1esv,1vesnc}) considering metrics
of type $g_{ij}\sim e^{\lambda (x,y)}\ ^{L}g_{ij}(x,y)\ $\ which can not be
derived directly from a mechanical Lagrangian. The ideas and methods to work
with arbitrary symmetric and nondegenerated tensor fields $g_{ij}(x,y)$ were
concluded in geometric and physical models for generalized Lagrange spaces,
denoted $GL^{n}=(M,g_{ij}(x,y)),$ on $\widetilde{TM},$ see \cite{1ma1987,1ma}%
, where $g_{ij}(x,y)$ is called the \textbf{fundamental tensor field}. Of
course, the geometric constructions will be equivalent if we shall work on
N--anholonomic manifolds $\mathbf{V}^{n+n}$ with nonholonomic coordinates $%
y. $ If we prescribe an arbitrary N--connection $N_{i}^{a}(x,y)$ and
consider that a metric $g_{ij}$ defines both the h-- and -v--components of a
d--metric (\ref{m1}), we can introduce the canonical d--connection (\ref%
{3cdctb}) and compute the components of d--curvature (\ref{3dcurvtb}),
define Ricci and Einstein tensors, elaborate generalized Lagrange models of
gravity.

If we work with a general fundamental tensor field $g_{ij}$ which can not be
transformed into $\ ^{L}g_{ij},$ we can consider an effective Lagrange
function \footnote{%
in \cite{1ma1987,1ma}, it is called the absolute energy of a $GL^{n}$%
--space, but for further applications in modern gravity the term ''energy''
may result in certain type ambiguities},
$\mathcal{L}(x,y)\doteqdot g_{ab}(x,y)y^{a}y^{b}$ and use
\begin{equation}
\ ^{\mathcal{L}}g_{ab}\doteqdot \frac{1}{2}\frac{\partial ^{2}\mathcal{L}}{%
\partial y^{a}\partial y^{b}}  \label{glhf}
\end{equation}%
as a Lagrange Hessian (\ref{slm}). A space $GL^{n}=(M,g_{ij}(x,y))$ is said
to be with a weakly regular metric if $L^{n}=\left[ M,L=\sqrt{|\mathcal{L}|})%
\right] $ is a Lagrange space. For such spaces, we can define a canonical
nonlinear connection structure%
\begin{equation}
\ ^{\mathcal{L}}N_{j}^{a}(x,y)\doteqdot \frac{\partial \ ^{\mathcal{L}}G^{a}%
}{\partial y^{j}},  \label{glnc}
\end{equation}%
\begin{equation*}
\ ^{\mathcal{L}}G^{a} =\frac{1}{4}\ ^{\mathcal{L}}g^{ab}\left( y^{k}\frac{%
\partial \mathcal{L}}{\partial y^{b}\partial x^{k}}-\frac{\partial \mathcal{L%
}}{\partial x^{a}}\right) = \frac{1}{4}\ ^{\mathcal{L}}g^{ab}\left( \frac{%
\partial g_{bc}}{\partial y^{d}}+\frac{\partial g_{bd}}{\partial y^{c}}-%
\frac{\partial g_{cd}}{\partial y^{b}}\right) y^{c}y^{d},
\end{equation*}
which allows us to write $\ ^{\mathcal{L}}N_{j}^{a}$ is terms of the
fundamental tensor field $g_{ij}(x,y).$ The geometry of such generalized
Lagrange spaces is completely similar to that of usual Lagrange one, with
that difference that we start not with a Lagrangian but with a fundamental
tensor field.

In our papers \cite{1vsh1,1vsh2}, we considered nonholonomic transforms of a
metric $\ ^{\mathcal{L}}g_{a^{\prime }b^{\prime }}(x,y)$
\begin{equation}
g_{ab}(x,y)=e_{a}^{\ a^{\prime }}(x,y)e_{b}^{\ b^{\prime }}(x,y)\ ^{\mathcal{%
L}}g_{a^{\prime }b^{\prime }}(x,y)  \label{ftgfs}
\end{equation}%
when $\ ^{\mathcal{L}}g_{a^{\prime }b^{\prime }}\doteqdot \frac{1}{2}\left(
e_{a^{\prime }}e_{b^{\prime }}\mathcal{L}+e_{b^{\prime }}e_{a^{\prime }}%
\mathcal{L}\right) =\ ^{0}g_{a^{\prime }b^{\prime }},$ for $e_{a^{\prime
}}=e_{\ a^{\prime }}^{a}(x,y)\frac{\partial }{\partial y^{a}},$ where $\
^{0}g_{a^{\prime }b^{\prime }}$ are constant coefficients (or in a more
general case, they should result in a constant matrix for the d--curvatures (%
\ref{dcurv}) of a canonical d--connection (\ref{candcon})). Such
constructions allowed to derive proper solitonic hierarchies and
bi--Hamilton structures for any (pseudo) Riemannian or generalized
Finsler--Lagrange metric. The point was to work not with the Levi Civita
connection (for which the solitonic equations became very cumbersome) but
with a correspondingly defined canonical d--connection allowing to apply
well defined methods from the geometry of nonlinear connections. Having
encoded the ''gravity and geometric mechanics'' information into solitonic
hierarchies and convenient d--connections, the constructions were shown to
hold true if they are ''inverted'' to those with usual Levi Civita
connections.

\subsection{An ansatz for constructing exact solutions}

\label{exsolans}We consider a four dimensional (4D) manifold $\mathbf{V}$ of
necessary smooth class and conventional splitting of dimensions $\dim
\mathbf{V=}$ $n+m$ for $n=2$ and $m=2.$ The local coordinates are labelled
in the form $u^{\alpha }=(x^{i},y^{a})=(x^{1},x^{2},y^{3}=v,y^{4}),$ for $%
i=1,2$ and $a,b,...=3,4.$

The ansatz of type (\ref{m1}) is parametrized in the form
\begin{eqnarray}
\mathbf{g} &=&g_{1}(x^{i}){dx^{1}}\otimes {dx^{1}}+g_{2}\left( x^{i}\right) {%
dx^{2}}\otimes {dx^{2}}  \notag \\
&& + h_{3}\left( x^{k},v\right) \ {\delta v}\otimes {\delta v}+h_{4}\left(
x^{k},v\right) \ {\delta y}\otimes {\delta y},  \notag \\
\delta v &=&dv+w_{i}\left( x^{k},v\right) dx^{i},\ \delta y=dy+n_{i}\left(
x^{k},v\right) dx^{i}  \label{ans5d}
\end{eqnarray}%
with the coefficients defined by some necessary smooth class functions of
type $%
g_{1,2}=g_{1,2}(x^{1},x^{2}),h_{3,4}=h_{3,4}(x^{i},v),w_{i}=w_{i}(x^{k},v),n_{i}=n_{i}(x^{k},v).
$ The off--diagonal terms of this metric, written with respect to the
coordinate dual frame $du^{\alpha }=(dx^{i},dy^{a}),$ can be redefined to
state a N--connection structure $\mathbf{N}=[N_{i}^{3}=w_{i}(x^{k},v),$$%
N_{i}^{4}=n_{i}(x^{k},v)]$ with a N--elongated co--frame (\ref{ddif})
parametrized as
\begin{equation}
e^{1}=dx^{1},\ e^{2}=dx^{2},\mathbf{e}^{3}=\delta v=dv+w_{i}dx^{i},\ \mathbf{%
e}^{4}=\delta y=dy+n_{i}dx^{i}.  \label{ddif5}
\end{equation}%
This vielbein is dual to the local basis%
\begin{equation}
\mathbf{e}_{i}=\frac{\partial }{\partial x^{i}}-w_{i}\left( x^{k},v\right)
\frac{\partial }{\partial v}-n_{i}\left( x^{k},v\right) \frac{\partial }{%
\partial y^{5}},e_{3}=\frac{\partial }{\partial v},e_{4}=\frac{\partial }{%
\partial y^{5}},  \label{dder5}
\end{equation}%
which is a particular case of the N--adapted frame (\ref{dder}). The metric (%
\ref{ans5d}) does not depend on variable $y^{4},$ i.e. it possesses a
Killing vector $e_{4}=\partial /\partial y^{4},$ and distinguish the
dependence on the so--called ''anisotropic'' variable $y^{3}=v.$

Computing the components of the Ricci and Einstein tensors for the metric (%
\ref{ans5d}) and canonical d--connection (see details on tensors components'
calculus in Refs. \cite{1valg,1vsgg}), one proves that the Einstein
equations (\ref{einstgrdef}) for a diagonal with respect to (\ref{ddif5})
and (\ref{dder5}) source,%
\begin{equation}
\mathbf{\Upsilon }_{\beta }^{\alpha }+\ ^{Z}\mathbf{\Upsilon }_{\beta
}^{\alpha }=[\Upsilon _{1}^{1}=\Upsilon _{2}(x^{i},v),\Upsilon
_{2}^{2}=\Upsilon _{2}(x^{i},v),\Upsilon _{3}^{3}=\Upsilon
_{4}(x^{i}),\Upsilon _{4}^{4}=\Upsilon _{4}(x^{i})]  \label{sdiag}
\end{equation}%
transform into this system of partial differential equations:
\begin{eqnarray}
\widehat{R}_{1}^{1} &=&\widehat{R}_{2}^{2}  \label{ep1a} \\
&=&\frac{1}{2g_{1}g_{2}}[\frac{g_{1}^{\bullet }g_{2}^{\bullet }}{2g_{1}}+%
\frac{(g_{2}^{\bullet })^{2}}{2g_{2}}-g_{2}^{\bullet \bullet }+\frac{%
g_{1}^{^{\prime }}g_{2}^{^{\prime }}}{2g_{2}}+\frac{(g_{1}^{^{\prime }})^{2}%
}{2g_{1}}-g_{1}^{^{\prime \prime }}]=-\Upsilon _{4}(x^{i}),  \notag \\
\widehat{S}_{3}^{3} &=&\widehat{S}_{4}^{4}=\frac{1}{2h_{3}h_{4}}\left[
h_{4}^{\ast }\left( \ln \sqrt{|h_{3}h_{4}|}\right) ^{\ast }-h_{4}^{\ast \ast
}\right] =-\Upsilon _{2}(x^{i},v),  \label{ep2a} \\
\widehat{R}_{3i} &=&-w_{i}\frac{\beta }{2h_{4}}-\frac{\alpha _{i}}{2h_{4}}=0,
\label{ep3a} \\
\widehat{R}_{4i} &=&-\frac{h_{3}}{2h_{4}}\left[ n_{i}^{\ast \ast }+\gamma
n_{i}^{\ast }\right] =0,  \label{ep4a}
\end{eqnarray}%
where, for $h_{3,4}^{\ast }\neq 0,$%
\begin{eqnarray}
\alpha _{i} &=&h_{4}^{\ast }\partial _{i}\phi ,\ \beta =h_{4}^{\ast }\ \phi
^{\ast },\ \gamma =\frac{3h_{4}^{\ast }}{2h_{4}}-\frac{h_{3}^{\ast }}{h_{3}},
\label{coef} \\
\phi &=&\ln |h_{3}^{\ast }/\sqrt{|h_{3}h_{4}|}|,  \label{coefa}
\end{eqnarray}%
when the necessary partial derivatives are written in the form \ $a^{\bullet
}=\partial a/\partial x^{1},$ $a^{\prime }=\partial a/\partial x^{2},$\ $%
a^{\ast }=\partial a/\partial v.$ In the vacuum case, we must consider $%
\Upsilon _{2,4}=0.$ We note that we use a source of type (\ref{sdiag}) in
order to show that the anholonomic frame method can be applied also for
non--vacuum solutions, for instance, when $\Upsilon _{2}=\lambda _{2}=const$
and $\Upsilon _{4}=\lambda _{4}=const,$ defining locally anisotropic
configurations generated by an anisotropic cosmological constant, which in
its turn, can be induced by certain ansatz for the so--called $H$--field
(absolutely antisymmetric third rank tensor field) in string theory \cite%
{1vesnc,1vsgg,1valg}. Here we note that the off--diagonal gravitational
interactions can model locally anisotropic configurations even if $\lambda
_{2}=\lambda _{4},$ or both values vanish.

In string gravity, the nontrivial torsion components and source $\kappa
\mathbf{\Upsilon }_{\alpha \beta }$ can be related to certain effective
interactions with the strength (torsion)
\begin{equation*}
H_{\mu \nu \rho }=\mathbf{e}_{\mu }B_{\nu \rho }+\mathbf{e}_{\rho }B_{\mu
\nu }+\mathbf{e}_{\nu }B_{\rho \mu }
\end{equation*}%
of an antisymmetric field $B_{\nu \rho },$ when%
\begin{eqnarray}
R_{\mu \nu }&=&-\frac{1}{4}H_{\mu }^{\ \nu \rho }H_{\nu \lambda \rho }
\label{c01}    \\
D_{\lambda }H^{\lambda \mu \nu } &=&0,  \label{c02}
\end{eqnarray}%
see details on string gravity, for instance, in Refs. \cite%
{1string1,1string2}. The conditions (\ref{c01}) and (\ref{c02}) are
satisfied by the ansatz
\begin{equation}
H_{\mu \nu \rho }=\widehat{Z}_{\mu \nu \rho }+\widehat{H}_{\mu \nu \rho
}=\lambda _{\lbrack H]}\sqrt{\mid g_{\alpha \beta }\mid }\varepsilon _{\nu
\lambda \rho }  \label{ansh}
\end{equation}%
where $\varepsilon _{\nu \lambda \rho }$ is completely antisymmetric and the
distorsion (from the Levi--Civita connection) and $\widehat{Z}_{\mu \alpha
\beta }\mathbf{e}^{\mu }=\mathbf{e}_{\beta }\rfloor \mathcal{T}_{\alpha }-%
\mathbf{e}_{\alpha }\rfloor \mathcal{T}_{\beta }+\frac{1}{2}\left( \mathbf{e}%
_{\alpha }\rfloor \mathbf{e}_{\beta }\rfloor \mathcal{T}_{\gamma }\right)
\mathbf{e}^{\gamma }$ is defined by the torsion tensor (\ref{tors}). Our $H$%
--field ansatz is different from those already used in string gravity when $%
\widehat{H}_{\mu \nu \rho }=\lambda _{\lbrack H]}\sqrt{\mid g_{\alpha \beta
}\mid }\varepsilon _{\nu \lambda \rho }.$ \ In our approach, we define $%
H_{\mu \nu \rho }$ and $\widehat{Z}_{\mu \nu \rho }$ from the respective
ansatz for the $H$--field and nonholonomically deformed metric, compute the
torsion tensor for the canonical distinguished connection and, finally,
define the 'deformed' H--field as $\widehat{H}_{\mu \nu \rho }=\lambda
_{\lbrack H]}\sqrt{\mid g_{\alpha \beta }\mid }\varepsilon _{\nu \lambda
\rho }-\widehat{Z}_{\mu \nu \rho }.$

Summarizing the results for an ansatz (\ref{ans5d}) with arbitrary
signatures $\epsilon _{\alpha }=\left( \epsilon _{1},\epsilon _{2},\epsilon
_{3},\epsilon _{4}\right) $ (where $\epsilon _{\alpha }=\pm 1)$ and $%
h_{3}^{\ast }\neq 0$ and $h_{4}^{\ast }\neq 0,$ one proves \cite%
{1vesnc,1valg,1vsgg} that any off--diagonal metric
\begin{eqnarray}
&&\ ^{\circ }\mathbf{g} =e^{\psi (x^{i})}\left[ \epsilon _{1}\ dx^{1}\otimes
dx^{1}+\epsilon _{2}\ dx^{2}\otimes dx^{2}\right] +\epsilon
_{3}h_{0}^{2}(x^{i}) \times  \notag \\
&& \left[ f^{\ast }\left( x^{i},v\right) \right] ^{2}|\varsigma \left(
x^{i},v\right) |\ \delta v\otimes \delta v +\epsilon _{4}\left[ f\left(
x^{i},v\right) -f_{0}(x^{i})\right] ^{2}\ \delta y^{4}\otimes \delta y^{4},
\notag \\
&&\delta v =dv+w_{k}\left( x^{i},v\right) dx^{k},\ \delta
y^{4}=dy^{4}+n_{k}\left( x^{i},v\right) dx^{k},  \label{gensol1}
\end{eqnarray}%
where $\psi (x^{i})$ is a solution of the 2D equation $\epsilon _{1}\psi
^{\bullet \bullet }+\epsilon _{2}\psi ^{^{\prime \prime }}=\Upsilon _{4},$
\begin{equation*}
\varsigma \left( x^{i},v\right) =\varsigma _{\lbrack 0]}\left( x^{i}\right) -%
\frac{\epsilon _{3}}{8}h_{0}^{2}(x^{i})\int \Upsilon _{2}(x^{k},v)f^{\ast
}\left( x^{i},v\right) \left[ f\left( x^{i},v\right) -f_{0}(x^{i})\right] dv,
\end{equation*}%
for a given source $\Upsilon _{4}\left( x^{i}\right) ,$
 and the N--connection coefficients $N_{i}^{3}=w_{i}(x^{k},v)$ and $%
N_{i}^{4}=n_{i}(x^{k},v)$ are computed following the formulas
\begin{eqnarray}
w_{i}&=&-\frac{\partial _{i}\varsigma \left( x^{k},v\right) }{\varsigma ^{\ast
}\left( x^{k},v\right) }  \label{gensol1w}   \\
n_{k}&=&\ ^{1}n_{k}\left( x^{i}\right) +\ ^{2}n_{k}\left( x^{i}\right) \int
\frac{\left[ f^{\ast }\left( x^{i},v\right) \right] ^{2}}{\left[ f\left(
x^{i},v\right) -f_{0}(x^{i})\right] ^{3}}\varsigma \left( x^{i},v\right) dv,
\label{gensol1n}
\end{eqnarray}%
define an exact solution of the system of Einstein equations (\ref{ep1a})--(%
\ref{ep4a}). It should be emphasized that such solutions depend on arbitrary
nontrivial functions $f\left( x^{i},v\right) $ (with $f^{\ast }\neq 0),$ $%
f_{0}(x^{i}),$ $h_{0}^{2}(x^{i})$, $\ \varsigma _{\lbrack 0]}\left(
x^{i}\right) ,$ $\ ^{1}n_{k}\left( x^{i}\right) $ and $\ \ ^{2}n_{k}\left(
x^{i}\right) ,$ and sources $\Upsilon _{2}(x^{k},v),\Upsilon _{4}\left(
x^{i}\right) .$ Such values for the corresponding signatures $\epsilon
_{\alpha }=\pm 1$ have to be defined by certain boundary conditions and
physical considerations. These classes of solutions depending on integration
functions are more general than those for diagonal ansatz depending, for
instance, on one radial variable like in the case of the Schwarzschild
solution (when the Einstein equations are reduced to an effective nonlinear
ordinary differential equation, ODE). In the case of ODE, the integral
varieties depend on integration constants which can be defined from certain
boundary/ asymptotic and symmetry conditions, for instance, from the
constraint that far away from the horizon the Schwarzschild metric contains
corrections from the Newton potential. Because the ansatz (\ref{ans5d})
results in a system of nonlinear partial differential equations (\ref{ep1a}%
)--(\ref{ep4a}), the solutions depend not only on integration constants, but
on very general classes of integration functions.

The ansatz of type (\ref{ans5d}) with $h_{3}^{\ast }=0$ but $h_{4}^{\ast
}\neq 0$ (or, inversely, $h_{3}^{\ast }\neq 0$ but $h_{4}^{\ast }=0)$
consist more special cases and request a bit different method of
constructing exact solutions. Nevertheless, such type solutions are also
generic off--diagonal and they may be of substantial interest (the length of
paper does not allow to include an analysis of such particular cases).

A very general class of exact solutions of the Einstein equations with
nontrivial sources (\ref{sdiag}), in general relativity, is defined by the
ansatz
\begin{eqnarray}
\ _{\shortmid }^{\circ }\mathbf{g} &=&e^{\psi (x^{i})}\left[ \epsilon _{1}\
dx^{1}\otimes dx^{1}+\epsilon _{2}\ dx^{2}\otimes dx^{2}\right]
\label{eeqsol} \\
&&+h_{3}\left( x^{i},v\right) \ \delta v\otimes \delta v+h_{4}\left(
x^{i},v\right) \ \delta y^{4}\otimes \delta y^{4},  \notag \\
\delta v &=&dv+w_{1}\left( x^{i},v\right) dx^{1}+w_{2}\left( x^{i},v\right)
dx^{2},  \notag \\
 \delta y^{4} &=& dy^{4}+n_{1}\left( x^{i}\right) dx^{1}+n_{2}\left(
x^{i}\right) dx^{2},  \notag
\end{eqnarray}%
with the coefficients restricted to satisfy the conditions
\begin{eqnarray}
\epsilon _{1}\psi ^{\bullet \bullet }+\epsilon _{2}\psi ^{^{\prime \prime }}
&=&\Upsilon _{4},\ h_{4}^{\ast }\phi /h_{3}h_{4} =\Upsilon _{2},
\label{ep2b} \\
w_{1}^{\prime }-w_{2}^{\bullet }+w_{2}w_{1}^{\ast }-w_{1}w_{2}^{\ast }
&=&0,\ n_{1}^{\prime }-n_{2}^{\bullet } = 0,  \notag
\end{eqnarray}%
for $w_{i}=\partial _{i}\phi /\phi ^{\ast },$ see (\ref{coefa}), for given
sources $\Upsilon _{4}(x^{k})$ and $\Upsilon _{2}(x^{k},v).$ We note that
the second equation in (\ref{ep2b}) relates two functions $h_{3}$ and $h_{4}$
and the third and forth equations satisfy the conditions (\ref{intcond}).

Even the ansatz (\ref{ans5d}) depends on three coordinates $(x^{k},v),$ it
allows us to construct more general classes of solutions for d--metrics,
depending on four coordinates: such solutions can be related by chains of
nonholonomic transforms. New classes of generic off--diagonal solutions will
describe nonhlonomic Einstein spaces related to string gravity, if one of
the chain metric is of type (\ref{gensol1}), or in Einstein gravity, if one
of the chain metric is of type (\ref{eeqsol}).

\section{Einstein Gravity and Lagrange--K\"{a}h\-ler Spaces}

\label{sakeg} We show how nonholonomic Riemannian spaces can be transformed
into almost Hermitian manifolds enabled with nonintegrable almost complex
structures.

\subsection{Almost Hermitian connections and general relativity}

We prove that the Einstein gravity on a (pseudo) Riemannian manifold $%
V^{n+n} $ can be equivalently redefined as an almost Hermitian model if a
nonintegrable N--connection splitting is prescribed. The Einstein theory can
be also modified by considering certain canonical lifts on tangent bundles.
The first class of Finsler--Lagrange like models \cite{1vqgr3} preserves the
local Lorentz symmetry and can be applied for constructing exact solutions
in Einstein gravity or for developing some approaches to quantum gravity
following methods of geometric/deformation quantization. The second class of
such models \cite{1vqgr2} can be considered for some extensions to canonical
quantum theories of gravity which can be elaborated in a renormalizable
form, but, in general, result in violation of local Lorentz symmetry by such
quantum effects.

\subsubsection{Nonholonomic deformations in Einstein gravity}

Let us consider a metric $\underline{g}_{\alpha \beta }$ (\ref{ansatz}),
which for a $\left( n+n\right) $--splitting by a set of prescribed
coefficients $N_{i}^{a}(x,y)$ can be represented as a d--metric $\mathbf{g}$
(\ref{m1}). Respectively, we can
write the Einstein equations in the form (\ref{einstgr}), or, equivalently,
in the form (\ref{einstgrdef}) with the source $\ ^{Z}\mathbf{\Upsilon }%
_{\alpha \beta }$ defined by the off--diagonal metric coefficients of $%
\underline{g}_{\alpha \beta },$ depending linearly on $N_{i}^{a},$ and
generating the distorsion tensor $\ _{\shortmid }Z_{\ \alpha \beta }^{\gamma
}.$

Computing the Ricci and Einstein d--tensors, we conclude that the Einstein
equations written in terms of the almost Hermitian d--connection can be also
parametrized in the form (\ref{einstgrdef}). Such geometric structures are
nonholonomic: working respectively with $\underline{g},$ $\mathbf{g},$ we
elaborate equivalent geometric and physical models on $V^{n+n},\mathbf{V}%
^{n+n}.$  Even for vacuum configurations, when $\mathbf{\Upsilon }%
_{\alpha \beta }=0,$ in the almost Hermitian model of the Einstein gravity,
we have an effective source $\ ^{Z}\mathbf{\Upsilon }_{\alpha \beta }$
induced by the coefficients of generic off--diagonal metric. Nevertheless,
there are possible integrable configurations, when the conditions (\ref%
{intcond}) are satisfied. In this case, $^{Z}\mathbf{\Upsilon }_{\alpha
\beta }=0,$ and we can construct effective Hermitian configurations defining
vacuum Einstein foliations.

One should be noted that the geometry of nonholonomic $2+2$ splitting in
general relativity, with nonholonomic frames and d--connections, or almost
Hermitian connections, is very different from the geometry of the well known
$3+1$ splitting ADM formalism, see \cite{1mtw}, when only the Levi Civita
connection is used. Following the anholonomic frame method, we work with
different classes of connections and frames when some new symmetries and
invariants are distinguished and the field equations became exactly
integrable for some general metric ansatz. Constraining or redefining the
integral varieties and geometric objects, we can generate, for instance,
exact solutions in Einstein gravity and compute quantum corrections to such
solutions.

\subsubsection{Conformal lifts of Einstein structures to tangent bund\-les}

Let us consider a pseudo--Riemannian manifold $M$ enabled with a metric $\
_{\shortmid }g_{ij}(x)$ as a solution of the Einstein equations. We define a
procedure lifting $_{\shortmid }g_{ij}(x)$ conformally on $TM$ and inducing
a generalized Lagrange structure and a corresponding almost Hermitian
geometry. Let us introduce
$\ ^{\varpi }\mathcal{L}(x,y)\doteqdot \varpi ^{2}(x,y)g_{ab}(x)y^{a}y^{b}$
 and use
\begin{equation}
\ ^{\varpi }g_{ab}\doteqdot \frac{1}{2}\frac{\partial ^{2}\ ^{\varpi }%
\mathcal{L}}{\partial y^{a}\partial y^{b}}  \label{cslm}
\end{equation}%
as a Lagrange Hessian for (\ref{slm}). A space $GL^{n}=(M,\ ^{\varpi
}g_{ij}(x,y))$ possess a weakly regular conformally deformed metric if $%
L^{n}=\left[ M,L=\sqrt{|\ ^{\varpi }\mathcal{L}|})\right] $ is a Lagrange
space. We can construct a canonical N--connection $\ ^{\varpi }N_{i}^{a}$
following formulas (\ref{glnc}), using $\ ^{\varpi }\mathcal{L}$ instead of $%
\mathcal{L}$ and $\ ^{\varpi }g_{ab}$ instead of $\ ^{\mathcal{L}}g_{ab}$ (%
\ref{glhf}), and define a d--metric on $TM,$%
\begin{equation}
\ ^{\varpi }\mathbf{g}=\ ^{\varpi }g_{ij}(x,y)\ dx^{i}\otimes dx^{j}+\
^{\varpi }g_{ij}(x,y)\ ^{\varpi }\mathbf{e}^{i}\otimes \ ^{\varpi }\mathbf{e}%
^{j},  \label{eslm}
\end{equation}%
where $^{\varpi }\mathbf{e}^{i}=dy^{i}+$ $^{\varpi }N_{i}^{j}dx^{i}.$ The
canonical d--connection and corresponding curvatures are constructed as in
generalized Lagrange geometry but using $\ ^{\varpi }\mathbf{g.}$

For the d--metric (\ref{eslm}), the model is elaborated for tangent bundles
with holonomic vertical frame structure. The linear operator $\mathbf{F}$
defining the almost complex structure acts on $\mathbf{TM}$ following
formulas $\mathbf{F}(\ ^{\varpi }\mathbf{e}_{i})=-\partial _{i}$ and $%
\mathbf{F}(\partial _{i})=\ ^{\varpi }\mathbf{e}_{i},$ when $\mathbf{F\circ
F=-I,}$ for $\mathbf{I}$ being the unity matrix. The operator $\mathbf{F}$
reduces to a complex structure if and only if the h--distribution is
integrable.

The metric $\ ^{\varpi }\mathbf{g}$ (\ref{eslm}) induces a 2--form
associated to $\mathbf{F}$ following formulas $\ ^{\varpi }\mathbf{\theta
(X,Y)}\doteqdot \ ^{\varpi }\mathbf{g}\left( \mathbf{FX,Y}\right) $ for any
d--vectors $\mathbf{X}$ and $\mathbf{Y.}$ In local form, we have
$\ ^{\varpi }\mathbf{\theta }=\ ^{\varpi }g_{ij}(x,y)dy^{i}\wedge dx^{j}.$
 The canonical d--connection $\ ^{\varpi }\widehat{\mathbf{D}},$ with
N--adapted coefficients $^{\varpi }\mathbf{\Gamma }_{\ \beta \gamma
}^{\alpha }=\left( \ ^{\varpi }\ \widehat{L}_{\ jk}^{i},\ \ ^{\varpi }%
\widehat{C}_{\ bc}^{a}\right) ,$ and corresponding d--curvature has to be
computed with $\check{e}_{b}^{\ \underline{b}}=\delta _{b}^{\ \underline{b}}$
and $\ ^{\varpi }g_{ij}$ used instead of $g_{ij}.$

The model of almost Hermitian gravity $H^{2n}(\mathbf{TM},\ ^{\varpi }%
\mathbf{g,F})$ can be applied in order to construct different extensions of
general relativity to geometric quantum models on tangent bundle \cite%
{1vqgr2}. Such models will result positively in violation of local Lorentz
symmetry, because the geometric objects depend on fiber variables $y^{a}.$
The quasi--classical corrections can be obtained in the approximation $%
\varpi \sim 1.$ We omit in this work consideration of quantum models, but
note that Finsler methods and almost K\"{a}hler geometry seem to be very
useful for such generalizations of Einstein gravity.

\section{Finsler--Lagrange Metrics in Einstein \& String Gravity}

\label{sflexso}We consider certain general conditions when Lagrange and
Finsler structures can be modelled as exact solutions in string and Einstein
gravity. Then, we analyze two explicit examples of exact solutions of the
Einstein equations modelling generalized Lagrange--Finsler geometries and
nonholonomic deformations of physically valuable equations in Einstein
gravity to such locally anisotropic configurations.

\subsection{Einstein spaces modelling generalized Finsler structures}

In this section, we outline the calculation leading from generalized
Lagrange and Finsler structures to exact solutions in gravity. Let us consider
\begin{equation}
\ ^{\varepsilon }\mathbf{\check{g}}=\ ^{\varepsilon }g_{i^{\prime }j^{\prime
}}(x^{k^{\prime }},y^{l^{\prime }})\left( e^{i^{\prime }}\otimes
e^{j^{\prime }}+\mathbf{\check{e}}^{i^{\prime }}\otimes \mathbf{\check{e}}%
^{j^{\prime }}\right) ,  \label{m2}
\end{equation}%
where $\ ^{\varepsilon }g_{i^{\prime }j^{\prime }}$ can be any metric
defined by nonholonomic transforms (\ref{ftgfs}) or a v--metric $\ ^{%
\mathcal{L}}g_{ij}$ (\ref{glhf}), $\ ^{L}g_{ij}$ (\ref{lqf}), or $\
^{F}g_{ij}$ (\ref{fhes}). The co--frame h-- and v--bases
\begin{equation*}
e^{i^{\prime }} =\ e_{\ i}^{i^{\prime }}(x,y)\ dx^{i},\ \mathbf{\check{e}}%
^{a^{\prime }} =\check{e}_{\ a}^{a^{\prime }}(x,y)\ \delta y^{a}=\check{e}%
_{\ a}^{a^{\prime }}\left( dy^{a}+\ _{\shortmid }N_{i}^{a}dx^{i}\right)
=e^{a^{\prime }}+\ _{\shortmid }\check{N}_{i^{\prime }}^{a}e^{i^{\prime }},
\end{equation*}
define $e^{a^{\prime }}=\check{e}_{\ a}^{a^{\prime }}dy^{a}$ and $\ _{\shortmid
}\check{N}_{i^{\prime }}^{a}=\check{e}_{\ a}^{a^{\prime }}\ _{\shortmid
}N_{i^{\prime }}^{a}e_{\ i}^{i^{\prime }},$ when
\begin{equation}
^{\varepsilon }g_{i^{\prime }j^{\prime }}=e_{i^{\prime }}^{\ i}e_{i^{\prime
}}^{\ i}\ \ _{\shortmid }g_{ij},\ \ _{\shortmid }h_{ab}=\ ^{\varepsilon
}g_{a^{\prime }b^{\prime }}\check{e}_{\ a}^{a^{\prime }}\check{e}_{\
b}^{b^{\prime }},\ \ _{\shortmid }N_{i}^{a}=\eta _{i}^{a}(x,y)\
^{\varepsilon }N_{i}^{a},  \label{nhdfes}
\end{equation}%
where we do not consider summation on indices for ''polarization'' functions
$\eta _{i}^{a}$ and $^{\varepsilon }N_{i}^{a}$ is a canonical connection
corresponding to $^{\varepsilon }g_{i^{\prime }j^{\prime }}.$

The d--metric (\ref{m2}) is equivalently transformed into the d--metric
\begin{eqnarray}
\ ^{\varepsilon }\mathbf{\check{g}} &=&\ _{\shortmid }g_{ij}(x)dx^{i}\otimes
dx^{j}+\ _{\shortmid }h_{ab}(x,y)\delta y^{a}\otimes \delta y^{a},
\label{m2a} \\
\delta y^{a} &=&dy^{a}+\ _{\shortmid }N_{i}^{a}(x,y)dx^{i},  \notag
\end{eqnarray}%
where the coefficients $\ _{\shortmid }g_{ij}(x),\ \ _{\shortmid
}h_{ab}(x,y) $ and $\ _{\shortmid }N_{i}^{a}(x,y)$ are constrained to be
defined by a class of exact solutions (\ref{gensol1}), in string gravity, or
(\ref{eeqsol}), in Einstein gravity. If it is possible to get the limit $%
\eta _{i}^{a}\rightarrow 1,$ we can say that an exact solution (\ref{m2a})
models exactly a respective (generalized) Lagrange, or Finsler,
configuration. We argue that we define a nonholonomic deformation of a
Finsler (Lagrange) space given by data $^{\varepsilon }g_{i^{\prime
}j^{\prime }}$ and $\ ^{\varepsilon }N_{i}^{a}$ as a class of exact
solutions of the Einstein equations given by data $\ _{\shortmid }g_{ij},\
_{\shortmid }h_{ab}$ and $\ _{\shortmid }N_{i}^{a},$ for any $\eta
_{i}^{a}(x,y)\neq 1.$ Such constructions are possible, if certain nontrivial
values of $e_{i^{\prime }}^{\ i},\check{e}_{\ a}^{a^{\prime }}$ and $\eta
_{i}^{a}$ can be algebraically defined from relations (\ref{nhdfes}) for any
defined sets of coefficients of the d--metric (\ref{m2}) and (\ref{m2a}).

Expressing a solution in the form (\ref{m2}), we can define the
corresponding almost Hermitian 1--form$\mathbf{\check{\theta}}=g_{i^{\prime
}j^{\prime }}(x,y)\check{e}^{j^{\prime }}\wedge e^{i^{\prime }},$ and
construct an almost Hermitian geometry characterizing this solution for $%
\mathbf{\check{F}}(\mathbf{e}_{i^{\prime }})=-\check{e}_{i^{\prime }}$ and $%
\mathbf{\check{F}}(\check{e}_{i^{\prime }})=\mathbf{e}_{i^{\prime }},$ when $%
\mathbf{e}_{i^{\prime }}=e_{i^{\prime }}^{\ i}\left( \frac{\partial }{%
\partial x^{i}}-\ _{\shortmid }N_{i}^{a}\frac{\partial }{\partial y^{a}}%
\right) =e_{i^{\prime }}-\ _{\shortmid }\check{N}_{i^{\prime }}^{a^{\prime }}%
\check{e}_{a^{\prime }}.$ This is convenient for further applications to
certain models of quantum gravity and geometry. For explicit constructions
of the solutions, it is more convenient to work with parametrizations of
type (\ref{m2a}).

Finally, in this section, we note that the general properties of integral
varieties of such classes of solutions are discussed in Refs. \cite%
{1vparsol,1vsgg}.

\subsection{Deformation of Einstein exact solutions into Lagrange--Finsler
metrics}

Let us consider a metric ansatz $\ _{\shortmid }g_{\alpha \beta }$ (\ref{m1}%
) with quadratic metric interval%
\begin{eqnarray}
ds^{2} &=&\ _{\shortmid }g_{1}(x^{1},x^{2})\left( dx^{1}\right) ^{2}+\
_{\shortmid }g_{2}(x^{1},x^{2})\left( dx^{2}\right) ^{2}  \label{exsolpr} \\
&&+\ _{\shortmid }h_{3}(x^{1},x^{2},v)\left[ dv+\ _{\shortmid
}w_{1}(x^{1},x^{2},v)dx^{1}+\ _{\shortmid }w_{2}(x^{1},x^{2},v)dx^{2}\right]
^{2}  \notag \\
&&+\ _{\shortmid }h_{4}(x^{1},x^{2},v)\left[ dy^{4}+\ _{\shortmid
}n_{1}(x^{1},x^{2},v)dx^{1}+\ _{\shortmid }n_{2}(x^{1},x^{2},v)dx^{2}\right]
^{2}  \notag
\end{eqnarray}%
defining an exact solution of the Einstein equations (\ref{einstgr}), for
the Levi--Civita connection, when the source $\Upsilon _{\alpha \beta }$ is
zero or defined by a cosmological constant. We parametrize the coordinates
in the form $u^{\alpha }=(x^{1},x^{2},y^{3}=v,y^{4})$ and the N--connection
coefficients as $\ _{\shortmid }N_{i}^{3}=\ _{\shortmid }w_{i}$ and $%
_{\shortmid }N_{i}^{4}=\ _{\shortmid }n_{i}.$

We nonholonomically deform the coefficients of the \textbf{primary}
d--metric (\ref{exsolpr}), similarly to (\ref{nhdfes}), when the \textbf{%
target} quadratic interval
\begin{eqnarray}
ds_{\eta }^{2} &=&g_{i}\left( dx^{i}\right) ^{2}+h_{a}\left(
dy^{a}+N_{i}^{a}dx^{i}\right) ^{2} =e_{\ i}^{i^{\prime }}\ e_{\
j}^{j^{\prime }}\ ^{\varepsilon }g_{i^{\prime }j^{\prime }}dx^{i}dx^{j}
\label{targ1} \\
&+& \check{e}_{\ a}^{a^{\prime }}\ \check{e}_{\ b}^{b^{\prime }}\
^{\varepsilon }g_{a^{\prime }b^{\prime }}\left( dy^{a}+\eta _{i}^{a}\
^{\varepsilon }N_{i}^{a}dx^{i}\right) \left( dy^{b}+\eta _{j}^{b}\
^{\varepsilon }N_{j}^{b}dx^{j}\right)  \notag
\end{eqnarray}%
can be equivalently parametrized in the form%
\begin{eqnarray}
ds_{\eta }^{2} &=&\eta _{j}\ _{\shortmid }g_{j}(x^{i})\left( dx^{j}\right)
^{2}  \label{targ1a} \\
&&+\eta _{3}(x^{i},v)\ _{\shortmid }h_{3}(x^{i},v)\left[ dv+\ ^{w}\eta
_{i}(x^{k},v)\ ^{\varepsilon }w_{i}(x^{k},v)dx^{i}\right] ^{2}  \notag \\
&&+\eta _{4}(x^{i},v)\ _{\shortmid }h_{4}(x^{i},v)\left[ dy^{4}+\ ^{n}\eta
_{i}(x^{k},v)\ \ ^{\varepsilon }n_{i}(x^{k},v)dx^{i}\right] ^{2},  \notag
\end{eqnarray}%
similarly to ansatz (\ref{ans5d}), and defines a solution of type (\ref%
{gensol1}) (with N--connecti\-on coefficients (\ref{gensol1w}) and (\ref%
{gensol1n})), for the canonical d--connection, or a solution of type (\ref%
{eeqsol}) with the coefficients subjected to solve the conditions (\ref{ep2b}%
).

The class of target metrics (\ref{targ1}) and (\ref{targ1a}) defining the
result of a nonholonomic deformation of the primary data $[\ _{\shortmid
}g_{i},\ _{\shortmid }h_{a},\ _{\shortmid }N_{i}^{b}]$ to a
Finsler--Lagrange configuration $[^{\varepsilon }g_{i^{\prime }j^{\prime
}},\ ^{\varepsilon }N_{j}^{b}]$ are parametrized by vales $e_{\
i}^{i^{\prime }},\check{e}_{\ a}^{a^{\prime }}$ and $\eta _{i}^{a}.$ These
values can be expressed in terms of some generation and integration
functions and the coefficients of the primary and Finsler like d--metrics
and N--connections in such a manner when a primary class of exact solutions
is transformed into a ''more general'' class of exact solutions. In a
particular case, we can search for solutions when the target metrics
transform into primary metrics under some infinitesimal limits.

In general form, the solutions of equations (\ref{nhesp}) transformed into
the system of partial differential equations (\ref{ep1a})--(\ref{ep4a}), for
the d--metrics (\ref{targ1}), equivalently (\ref{targ1a}), are given by
corresponding sets of frame coefficients
\begin{eqnarray}
e_{\ 1}^{1^{\prime }} &=&\sqrt{|\eta _{1}|}\ \sqrt{|\ _{\shortmid }g_{1}|}%
\times \ ^{\varepsilon }E_{+},~e_{\ 1}^{2^{\prime }}=\sqrt{|\eta _{2}|}\
\sqrt{|\ _{\shortmid }g_{1}|}\ /\ ^{\varepsilon }E_{+},  \notag \\
e_{\ 2}^{1^{\prime }} &=&-\sqrt{|\eta _{2}|}\ \sqrt{|\ _{\shortmid }g_{2}|}%
\times g_{1^{\prime }2^{\prime }}/\ ^{\varepsilon }E_{-},~e_{\ 2}^{2^{\prime
}}=\sqrt{|\eta _{2}|}\ \sqrt{|\ _{\shortmid }g_{2}|}\times \ ^{\varepsilon
}E_{-},  \label{hfr}  \\
e_{\ 3}^{3^{\prime }} &=&\sqrt{|\eta _{3}|}\ \sqrt{|\ _{\shortmid }h_{3}|}%
\times \ ^{\varepsilon }E_{+},~e_{\ 3}^{4^{\prime }}=\sqrt{|\eta _{3}|}\
\sqrt{|\ _{\shortmid }h_{3}|}\ /\ ^{\varepsilon }E_{+},  \notag \\
e_{\ 4}^{3^{\prime }} &=&-\sqrt{|\eta _{4}|}\ \sqrt{|\ _{\shortmid }h_{4}|}%
\times g_{1^{\prime }2^{\prime }}/\ ^{\varepsilon }E_{-},~e_{\ 4}^{4^{\prime
}}=\sqrt{|\eta _{4}|}\ \sqrt{|\ _{\shortmid }h_{4}|}\times \ ^{\varepsilon
}E_{-},  \label{vfr}
\end{eqnarray}%
where $\ ^{\varepsilon }E_{\pm }=\sqrt{|\ ^{\varepsilon }g_{1^{\prime
}1^{\prime }}\ ^{\varepsilon }g_{2^{\prime }2^{\prime }}\left[ \left( \
^{\varepsilon }g_{1^{\prime }1^{\prime }}\right) ^{2}\ ^{\varepsilon
}g_{2^{\prime }2^{\prime }}\pm \left( \ ^{\varepsilon }g_{1^{\prime
}2^{\prime }}\right) ^{3}\right] ^{-1}|}$ and h--polari\-za\-tions $\eta _{j}$
are defined from $g_{j}=\eta _{j}\ _{\shortmid }g_{j}(x^{i})=$ $\epsilon
_{j}e^{\psi (x^{i})},$ with signatures $\epsilon _{i}=\pm 1,$ for $\psi
(x^{i})$ being a solution of the 2D equation
\begin{equation}
\epsilon _{1}\psi ^{\bullet \bullet }+\epsilon _{2}\psi ^{^{\prime \prime
}}=\lambda ,  \label{ep2b1}
\end{equation}%
for a given source $\Upsilon _{4}\left( x^{i}\right) =\lambda ,$ and the
v--polarizations $\eta _{a}$ defined from the data $h_{a}=\eta _{a}\
_{\shortmid }h_{a},$ for
\begin{eqnarray}
 && h_{3}=\epsilon _{3}h_{0}^{2}(x^{i})\left[ f^{\ast }\left( x^{i},v\right) %
\right] ^{2}|\ ^{\lambda }\varsigma \left( x^{i},v\right) |,~h_{4}=\epsilon
_{4}\left[ f\left( x^{i},v\right) -f_{0}(x^{i})\right] ^{2},\nonumber
\\
&&\ ^{\lambda }\varsigma =\varsigma _{\lbrack 0]}\left(
x^{i}\right) -\frac{\epsilon _{3}}{8}\lambda h_{0}^{2}(x^{i})\int f^{\ast
}\left( x^{i},v\right) \left[ f\left( x^{i},v\right) -f_{0}(x^{i})\right] dv,
\label{ep2b2}
\end{eqnarray}%
for $\Upsilon _{2}(x^{k},v)=\lambda .$ The polarizations $\eta _{i}^{a}$ of
N--connection coefficients
$N_{i}^{3}=w_{i}=\ ^{w}\eta _{i}(x^{k},v)\ ^{\varepsilon }w_{i}(x^{k},v),\
N_{i}^{4}=n_{i}=\ ^{n}\eta _{i}(x^{k},v)\ \ ^{\varepsilon }n_{i}(x^{k},v)$
are computed from  respective formulas
\begin{eqnarray}
\ ^{w}\eta _{i}\ ^{\varepsilon }w_{i}&=&-\frac{\partial _{i}\ ^{\lambda
}\varsigma \left( x^{k},v\right) }{\ ^{\lambda }\varsigma ^{\ast }\left(
x^{k},v\right) },  \label{gensol1wl}  \\
\ ^{n}\eta _{k}\ ^{\varepsilon }n_{k}&=&\ ^{1}n_{k}\left( x^{i}\right) +\
^{2}n_{k}\left( x^{i}\right) \int \frac{\left[ f^{\ast }\left(
x^{i},v\right) \right] ^{2}\ ^{\lambda }\varsigma \left( x^{i},v\right)}{\left[ f\left( x^{i},v\right) -f_{0}(x^{i})%
\right] ^{3}} dv.
\label{gensol1nl}
\end{eqnarray}

We generate a class of exact solutions for Einstein spaces with $\Upsilon
_{2}=\Upsilon _{4}=\lambda $ if the integral varieties defined by $%
g_{j},h_{a},w_{i}$ and $n_{i}$ are subjected to constraints (\ref{ep2b}).

\subsection{Solitonic pp--waves and their effective Lagrange spaces}

Let us consider a d--metric of type (\ref{exsolpr}),
\begin{equation}
\delta s_{[pw]}^{2}=-dx^{2}-dy^{2}-2\kappa (x,y,v)\ dv^{2}+\ dp^{2}/8\kappa
(x,y,v),  \label{5aux5}
\end{equation}%
where the local coordinates are $\ x^{1}=x,\ x^{2}=y,\ y^{3}=v,\ y^{4}=p,$
and the nontrivial metric coefficients are parametrized%
$\ _{\shortmid }g_{1}=-1,\ \ _{\shortmid }g_{2}=-1,\ _{\shortmid
}h_{3}=-2\kappa (x,y,v),\ \ _{\shortmid }h_{4}=1/\ 8\ \kappa (x,y,v).$
This is vacuum solution of the Einstein equation defining pp--waves \cite%
{1peres}: for any $\kappa (x,y,v)$ solving $\kappa _{xx}+\kappa _{yy}=0,$%
with $v=z+t$ and $p=z-t,$ where $(x,y,z)$ are usual Cartesian coordinates
and $t$ is the time like coordinate. Two explicit examples of such solutions
are given by $ \kappa =(x^{2}-y^{2})\sin v,$
defining a plane monochromatic wave, or by
$\kappa ={xy} / {\left( x^{2}+y^{2}\right) ^{2}\exp \left[ v_{0}^{2}-v^{2}%
\right] },$  for $|v|<v_{0},$ and $\kappa = 0,$ for $|v|\geq v_{0},$
defining a wave packet travelling with unit velocity in the negative $z$
direction.

We nonholonomically deform the vacuum solution (\ref{5aux5}) to a d--metric
of type (\ref{targ1a})
\begin{eqnarray}
ds_{\eta }^{2} &=&-e^{\psi (x,y)}\left[ \left( dx\right) ^{2}+(dy)^{2}\right]
\label{5auxd} \\
&&-\eta _{3}(x,y,v)\ \cdot 2\kappa (x,y,v)\left[ dv+\ ^{w}\eta
_{i}(x,y,v,p)\ ^{\varepsilon }w_{i}(x,y,v,p)dx^{i}\right] ^{2}  \notag \\
&&+\eta _{4}(x,y,v)\ \cdot \frac{1}{8\kappa (x,y,v)}\left[ dy^{4}+\ ^{n}\eta
_{i}(x,y,v,p)\ \ ^{\varepsilon }n_{i}(x,y,v,p)dx^{i}\right] ^{2},  \notag
\end{eqnarray}%
where the polarization functions $\eta _{1}=\eta _{2}=e^{\psi (x,y)},\eta
_{3,4}(x,y,v),\ ^{w}\eta _{i}(x,y,v)$ \ and $\ ^{n}\eta _{i}(x,y,v)\ $\ have
to be defined as solutions in the form (\ref{ep2b1}), (\ref{ep2b2}), (\ref%
{gensol1wl}) and (\ref{gensol1nl}) for a string gravity ansatz (\ref{ansh}),
$\lambda =\lambda _{H}^{2}/2,$ \ and \ a prescribed (in this section)
analogous mechanical system with $\ $%
\begin{equation}
N_{i}^{a}=\{w_{i}(x,y,v)=\ ^{w}\eta _{i}\ ^{L}w_{i},n_{i}(x,y,v)=\ ^{n}\eta
_{i}\ ^{\varepsilon }n_{i}\}  \label{cnclnd}
\end{equation}%
for $\varepsilon =L(x,y,v,p)$ considered as regular Lagrangian modelled on a
N--anholonomic manifold with holonomic coordinates $(x,y)$ and nonholonomic
coordinates $(v,p).$

A class of 3D solitonic configurations can defined by taking a polarization
function $\eta _{4}(x,y,v)=\eta (x,y,v)$ as a solution of solitonic equation%
\footnote{%
as a matter of principle we can consider\ that $\eta $ is a solution of any
3D solitonic, or other, nonlinear wave equation.}
\begin{equation}
\eta ^{\bullet \bullet }+\epsilon (\eta ^{\prime }+6\eta \ \eta ^{\ast
}+\eta ^{\ast \ast \ast })^{\ast }=0,\ \epsilon =\pm 1,  \label{5solit1}
\end{equation}%
and $\eta _{1}=\eta _{2}=e^{\psi (x,y)}$ as a solution of (\ref{ep2b1})
written as
\begin{equation}
\psi ^{\bullet \bullet }+\psi ^{\prime \prime }=\frac{\lambda _{H}^{2}}{2}.
\label{5lapl}
\end{equation}%
Introducing the\ above stated data for the ansatz (\ref{5auxd}) into the
equation (\ref{ep2a}), we get two equations relating $h_{3}=\eta _{3}\
_{\shortmid }h_{3}$ and $h_{4}=\eta _{4}\ _{\shortmid }h_{4},$
\begin{eqnarray}
\eta _{4}&=&8\ \kappa (x,y,v)\left[ h_{4}^{[0]}(x,y)+\frac{1}{2\lambda _{H}^{2}%
}e^{2\eta (x,y,v)}\right]  \label{5sol2h5} \\
|\eta _{3}| &=& \frac{e^{-2\eta (x,y,v)}}{2\kappa ^{2}(x,y,v)}\left[
\left( \sqrt{|\eta _{4}(x,y,v)|}\right) ^{\ast }\right] ^{2},
\label{5sol2h4}
\end{eqnarray}%
where $h_{4}^{[0]}(x,y)$ is an integration function.

Having defined the coefficients $h_{a},$ we can solve the equations (\ref%
{ep3a}) and (\ref{ep4a}) expressing the coefficients (\ref{coef}) and (\ref%
{coefa}) through $\eta _{3}$ and $\eta _{4}$ defined by pp-- and solitonic
waves as in (\ref{5sol2h4}) and (\ref{5sol2h5}). The corresponding solutions
\begin{eqnarray}
w_{1}&=&\ ^{w}\eta _{1}\ ^{L}w_{1}=\left( \phi ^{\ast }\right) ^{-1}\partial
_{x}\phi ,\ w_{2}=\ ^{w}\eta _{1}\ ^{L}w_{1}=\left( \phi ^{\ast }\right)
^{-1}\partial _{y}\phi ,  \label{5sol2w} \\
n_{i}&=&n_{i}^{[0]}(x,y)+n_{i}^{[1]}(x,y)\int \left| \eta
_{3}(x,y,v)\eta _{4}^{-3/2}(x,y,v)\right| dv,  \label{5sol2na}
\end{eqnarray}
are for $\phi ^{\ast }=\partial \phi /\partial v,$ see formulas (\ref{coefa}),
where $n_{i}^{[0]}(x,y)$ and $n_{i}^{[1]}(x,y)$ are integration functions.
The values $e^{\psi (x,y)},$ $\eta _{3}$ (\ref{5sol2h4}), $\eta _{4}$ (\ref%
{5sol2h5}), $w_{i}$ (\ref{5sol2w}) and $n_{i}$ (\ref{5sol2na}) for the
ansatz (\ref{5auxd}) completely define a nonlinear superpositions of
solitonic and pp--waves as an exact solution of the Einstein equations in
string gravity if there are prescribed some initial values for the nonlinear
waves under consideration. In general, such solutions depend on some classes
of generation and integration functions.

It is possible to give a regular Lagrange analogous interpretation of an
explicit exact solution (\ref{5auxd}) if we prescribe a regular Lagrangian $%
\varepsilon =L(x,y,v,p),$ with Hessian $\ ^{L}g_{i^{\prime }j^{\prime }}=%
\frac{1}{2}\frac{\partial ^{2}L}{\partial y^{i^{\prime }}\partial
y^{j^{\prime }}},$ for $x^{i^{\prime }}=(x,y)$ and $y^{a^{\prime }}=(v,p).$
Introducing the values $\ ^{L}g_{i^{\prime }j^{\prime }},$ $\eta _{1}=\eta
_{2}=e^{\psi },\eta _{3},\eta _{4}$ and $_{\shortmid }h_{3},\ \ _{\shortmid
}h_{4},$ all defined above, into (\ref{hfr}) and (\ref{vfr}), we compute the
vierbein coefficients $e_{\ i}^{i^{\prime }}$ and $\check{e}_{\
a}^{a^{\prime }}$ which allows us to redefine equivalently the quadratic
element in the form (\ref{targ1}),  for which
the N--connection coefficients $\ ^{L}N_{i}^{a}$ (\ref{cncl}) are
nonholonomically deformed to $N_{i}^{a}$ (\ref{cnclnd}). With respect to
such nonholonomic frames of references, an observer ''swimming in a string
gravitational ocean of interacting solitonic and pp--waves'' will see his
world as an analogous mechanical model defined by a regular Lagrangian $L.$

\subsection{Finsler--solitonic pp--waves in Schwarzschild spaces}

We consider a primary quadratic element
\begin{equation}
\delta s_{0}^{2}=-d\xi ^{2}-r^{2}(\xi )\ d\vartheta ^{2}-r^{2}(\xi )\sin
^{2}\vartheta \ d\varphi ^{2}+\varpi ^{2}(\xi )\ dt^{2},  \label{5aux1}
\end{equation}%
where the  nontrivial metric coefficients are
\begin{equation}
\ _{\shortmid }g_{2}=-r^{2}(\xi ),\ \ _{\shortmid }h_{3}=-r^{2}(\xi )\sin
^{2}\vartheta ,\ \ _{\shortmid }h_{4}=\varpi ^{2}(\xi ),  \label{5aux1p}
\end{equation}%
with $x^{1}=\xi ,x^{2}=\vartheta ,y^{3}=\varphi ,y^{4}=t,\ _{\shortmid }g_{1}=-1,$\
 $\xi =\int dr\ \left| 1-\frac{2\mu }{r}\right| ^{1/2}$ and $\varpi
^{2}(r)=1-\frac{2\mu }{r}.$ For $\mu $ being a point mass, the element (\ref%
{5aux1}) defines the Schwarzschild solution written in spacetime spherical
coordinates $(r,\vartheta ,\varphi ,t).$

Our aim, is to find a nonholonomic deformation of metric (\ref{5aux1}) to a
class of new vacuum solutions modelled by certain types of Finsler
geometries.

The target stationary metrics are parametrized in the form similar to (\ref%
{targ1a}), see also (\ref{eeqsol}),
\begin{eqnarray}
ds_{\eta }^{2} &=& -e^{\psi (\xi ,\vartheta )}\left[ \left( d\xi \right)
^{2}+r^{2}(\xi )(d\vartheta )^{2}\right] \label{5aux1pta} \\ &-& \eta _{3}(\xi ,\vartheta ,\varphi
)\ \cdot r^{2}(\xi )\sin ^{2}\vartheta \ [d\varphi +
  \ ^{w}\eta _{i}(\xi ,\vartheta ,\varphi ,t)\ ^{F}w_{i}(\xi ,\vartheta
,\varphi ,t)dx^{i}]^{2}  \notag \\ &+& \eta _{4}(\xi ,\vartheta ,\varphi )\ \cdot \varpi
^{2}(\xi )\ \left[ dt+\ ^{n}\eta _{i}(\xi ,\vartheta ,\varphi ,t)\
^{F}n_{i}(\xi ,\vartheta ,\varphi ,t)dx^{i}\right] ^{2}.  \notag
\end{eqnarray}%
The polarization functions $\eta _{1}=\eta _{2}=e^{\psi (\xi ,\vartheta
)},\eta _{a}(\xi ,\vartheta ,\varphi ),\ ^{w}\eta _{i}(\xi ,\vartheta
,\varphi )$ \ and $\ ^{n}\eta _{i}(\xi ,\vartheta ,\varphi )\ $\ have to be
defined as solutions of (\ref{ep2b}) for $\Upsilon _{2}=\Upsilon _{4}=0$ and
a prescribed (in this section) locally anisotropic, on $\varphi ,$ geometry
with $N_{i}^{a}=\{w_{i}(\xi ,\vartheta ,\varphi )=\ ^{w}\eta _{i}\
^{F}w_{i},\ n_{i}(\xi ,\vartheta ,\varphi )=\ ^{n}\eta _{i}\ ^{F}n_{i}\},$
for $\varepsilon =F^{2}(\xi ,\vartheta ,\varphi ,t)$ considered as a
fundamental Finsler function for a Finsler geometry modelled on a
N--anholonomic manifold with holonomic coordinates $(r,\vartheta )$ and
nonholonomic coordinates $(\varphi ,t).$ We note that even the values $\
^{w}\eta _{i},\ ^{F}w_{i},$ $\ ^{n}\eta _{i},$ and $\ ^{F}n_{i}$ can depend
on time like variable $t,$ such dependencies must result in N--connection
coefficients of type $N_{i}^{a}(\xi ,\vartheta ,\varphi ).$

Putting together the coefficients solving the Einstein equations (\ref{ep2a}%
)--(\ref{ep4a}) and (\ref{ep2b}), the class of vacuum solutions in general
relativity related to (\ref{5aux1pta}) can be parametrized in the form
\begin{eqnarray}
ds_{\eta }^{2} &=&-e^{\psi (\xi ,\vartheta )}\left[ \left( d\xi \right)
^{2}+r^{2}(\xi )(d\vartheta )^{2}\right]  \label{aux6} \\    &-&\ h_{0}^{2}\ \left[ b^{\ast }(\xi
,\vartheta ,\varphi )\right] ^{2}
 \left[ d\varphi +w_{1}(\xi ,\vartheta )d\xi +w_{2}(\xi ,\vartheta
)d\vartheta \right] ^{2}   \notag   \\
&+& \left[ b(\xi ,\vartheta ,\varphi )-b_{0}(\xi
,\vartheta )\right] ^{2}\ \left[ dt+n_{1}(\xi ,\vartheta )d\xi +n_{2}(\xi
,\vartheta )d\vartheta \right] ^{2},  \notag
\end{eqnarray}%
where $h_{0}=const$ and the coefficients are constrained to solve the
equations
\begin{equation}
\psi ^{\bullet \bullet }+\psi ^{^{\prime \prime }}=0,~w_{1}^{\prime
}-w_{2}^{\bullet }+w_{2}w_{1}^{\ast }-w_{1}w_{2}^{\ast }=0,~n_{1}^{\prime
}-n_{2}^{\bullet }=0,  \label{ep2bb}
\end{equation}%
for instance, for $w_{1}=(b^{\ast })^{-1}(b+b_{0})^{\bullet },$ $%
w_{2}=(b^{\ast })^{-1}(b+b_{0})^{\prime },n_{2}^{\bullet }=\partial
n_{2}/\partial \xi $ and $n_{1}^{^{\prime }}=\partial n_{1}/\partial
\vartheta .$

The polarization functions relating (\ref{aux6}) to (\ref{5aux1pta}), are
computed
\begin{eqnarray}
\eta _{1} &=&\eta _{2}=e^{\psi (\xi ,\vartheta )},\ \eta _{3}=\left[
h_{0}b^{\ast }/r(\xi )\sin \vartheta \right] ^{2},\eta _{4}=\left[
(b-b_{0})/\varpi \right] ^{2},  \label{polf6} \\
\ ^{w}\eta _{i} &=&w_{i}(\xi ,\vartheta )/\ ^{F}w_{i}(\xi ,\vartheta
,\varphi ,t),\ \ ^{n}\eta _{i}=n_{i}(\xi ,\vartheta )/\ ^{F}n_{i}(\xi
,\vartheta ,\varphi ,t).  \notag
\end{eqnarray}

The next step is to chose a Finsler geometry which will model (\ref{aux6}),
equivalently (\ref{5aux1pta}), as a Finsler like d--metric (\ref{targ1}).
For a fundamental Finsler function $F=F(\xi ,\vartheta ,\varphi ,t),$ where $%
x^{i^{\prime }}=(\xi ,\vartheta )$ are h--coordinates and $y^{a^{\prime
}}=\left( \varphi ,t\right) $ are v--coordinates, we compute $\
^{F}g_{a^{\prime }b^{\prime }}=(1/2)\partial ^{2}F/\partial y^{a^{\prime
}}\partial y^{b^{\prime }}$ following formulas (\ref{fhes}) and parametrize
the Cartan N--connection as $\ ^{C}N_{i}^{a}=\{\ ^{F}w_{i},\ ^{F}n_{i}\}.$
Introducing the values (\ref{5aux1p}),$\ ^{F}g_{i^{\prime }j^{\prime }}$ and
polarization functions (\ref{polf6}) into (\ref{hfr}) and (\ref{vfr}), we
compute the vierbein coefficients $e_{\ i}^{i^{\prime }}$ and $\check{e}_{\
a}^{a^{\prime }}$ which allows us to redefine equivalently the quadratic
element in the form (\ref{targ1}), in this case, for a Finsler configuration
for which the N--connection coefficients $\ ^{C}N_{i}^{a}$ (\ref{cncl}) are
nonholonomically deformed to $N_{i}^{a}$ satisfying the last two conditions
in (\ref{ep2bb}). With respect to such nonholonomic frames of references, an
observer ''swimming in a locally anisotropic gravitational ocean'' will see
the nonholonomically deformed Schwarzschild geometry as an analogous Finsler
model defined by a fundamental Finser function $F.$

\section{Outlook and Conclusions}

In this review article, we gave a self--contained account of the core
developments on generalized Finsler--Lagrange geometries and their modelling
on (pseudo) Riemannian and Riemann--Cartan manifolds provided with preferred
nonholonomic frame structure. We have shown how the Einstein gravity and
certain string models of gravity with torsion can be equivalently
reformulated in the language of generalized Finsler and almost Hermitian/ K%
\"{a}hler geometries. It was also argued that former criticism and
conclusions on experimental constraints and theoretical difficulties of
Finsler like gravity theories were grounded only for certain classes of
theories with metric noncompatible connections on tangent bundles and/or
resulting in violation of local Lorentz symmetry. We emphasized that there
were omitted the results when for some well defined classes of nonholonomic
transforms of geometric structures we can model geometric structures with
local anisotropy, of Finsler--Lagrange type, and generalizations, on
(pseudo) Riemann spaces and Einstein manifolds.

Our idea was to consider not only some convenient coordinate and frame
transforms, which simplify the procedure of constructing exact solutions,
but also to define alternatively new classes of connections which can be
employed to generate new solutions in gravity. We proved that the solutions
for the so--called canonical distinguished connections can be equivalently
re--defined for the Levi Civita connection and/ or constrained to define
integral varieties of solutions in general relativity.

The main conclusion of this work is that we can avoid all existing
experimental restrictions and theoretical difficulties of Finsler physical
models if we work with metric compatible Finsler like structures on
nonholonomic (Riemann, or Riemann--Cartan) manifolds but not on tangent
bundles. In such cases, all nonholonomic constructions modelled as exact
solutions of the Einstein and matter field equations (with various string,
quantum field ... corrections) are compatible with the standard paradigm in
modern physics.

In other turn, we emphasize that in quantum gravity, statistical and
thermodynamical models with local anisotropy, gauge theories with
constraints and broken symmetry and in geometric mechanics, nonholonomic
configurations on (co) tangent bundles, of Finsler type and generalizations,
metric compatible or with nonmetricity, seem to be also very important.

Various directions in generalized Finsler geometry and applications has
matured enough so that some tenths of monographs have been written,
including some recent and updated: we cite here \cite%
{1mats,1ma1987,1ma,1mhl,1mhf,1mhss,1mhh,1bej,1vsgg,1vmon1,1vstav,1bcs,1shen,1ant,1amaz,1as1,1as2,1ap,1bog}%
. These monographs approach and present the subjects from different
perspectives depending, of course, on the authors own taste, historical
period and interests both in geometry and physics. The monograph \cite{1vsgg}
summarizes and develops the results oriented to application of Finsler
methods to standard theories of gravity (on nonholonomic manifolds, not only
on tangent bundles) and their noncommutative generalizations; it was also
provided a critical analysis of the constructions with nonmetricity and
violations of local Lorentz symmetry.

Finally, we suggest the reader to see a more complete review \cite{1vrfg}
discussing  applications of Finsler and Lagrange geometry both to standard and
nonstandard models of physics (presenting a variant which
was not possible to be published because of limit of space in this journal).

\vskip5pt

\textbf{Acknowledgement: } The work is performed during a visit at the
Fields Institute. The author thanks M. Anastasiei, A. Bejancu, V. Ob\v{a}%
deanu and V. Oproiu for discussions and providing very important references
on the geometry of Finsler--Lagrange spaces, nonholonomic manifolds and
related almost K\"{a}hler geometry.
\end{document}